\documentclass[aps,pra,twocolumn,showpacs,superscriptaddress,floatfix]{revtex4-1}  % for review and submission
\usepackage{amsmath}
\usepackage{amsfonts}
\usepackage{amssymb}
\usepackage{braket}
\usepackage{mathtools}
\usepackage{epsfig}
\usepackage{graphicx}
\usepackage{color}
\usepackage{tikz}

\definecolor{darkgreen}{RGB}{0,100,20}

\begin{document}
\title{Mixed parity pairing in a dipolar gas}
\author{G.\ M.\ Bruun}
\affiliation{Department of Physics and Astronomy, University of Aarhus, Ny Munkegade, DK-8000 Aarhus C, Denmark}
\author{C.\ Hainzl and M.\ Laux}
\affiliation{Mathematisches Institut, Universit\"at T\"ubingen, Auf der Morgenstelle 10, 72076 T\"ubingen, Germany}

\date{\today}
\begin{abstract}
We show that fermionic dipoles in a two-layer geometry form Cooper pairs with both singlet and triplet components, when they 
are tilted with respect to the normal of the planes. The mixed parity 
pairing arises because the interaction between dipoles in the two different layers is not inversion symmetric. We use an efficient
 eigenvalue approach to calculate the zero temperature phase diagram of the system as a function of the dipole orientation and the layer distance. 
The phase diagram contains  purely triplet as well as mixed singlet and triplet superfluid phases. We show in detail  how the pair wave function for dipoles 
residing in different layers smoothly changes from singlet to triplet symmetry as the orientation of the dipoles is changed. Our results indicate that 
dipolar quantum gases can be used to unambiguously observe mixed parity pairing. 
\end{abstract}

\maketitle

\section{Introduction}
 Time-reversal symmetry and inversion symmetry play a key role in the formation of Cooper pairs. Anderson showed 
that time-reversal symmetry ensures the existence of degenerate states for $s$-wave singlet Cooper pairing~\cite{Anderson1959},
 whereas triplet pairing relies on inversion symmetry~\cite{Andersen1984}. 
The breaking of these symmetries has profound effects on pairing. Superconductivity 
in crystals with no inversion symmetry is intensely studied, since it
 can lead to Cooper pairs with both spin-singlet and spin-triplet components
and therefore no definite parity~\cite{Gorkov2001,Hayashi2006}. 
While there are many crystals lacking inversion symmetry, only few  experiments indicate mixed parity pairing.
 Evidence for  mixed parity pairing has been reported for 
  Li$_2$Pt$_3$B~\cite{Nishiyama2007,Yuan2006}, and measurements on CePt$_3$Si crystals  are 
consistent with mixed parity pairing~\cite{Bonalde2005, Izawa2005}, but can also be attributed to 
 multiband and disorder effects~\cite{Hidekazu2009}. 

In this article, we show that dipolar fermions residing in two parallel layers can form a superfluid characterised by mixed parity pairing. The 
dipoles are aligned by an external field, and when their dipole moment is tilted with respect to the normal of the plane, the interaction 
between dipoles residing in the two different layers is not inversion symmetric, which leads to mixed parity pairing with both singlet and triplet components. 
Using a computationally efficient eigenvalue approach, we investigate the competition between the mixed parity interlayer pairing  
and the pairing between dipoles in the same layers, which is  in the triplet channel. We map out the resulting zero temperature phase diagram as a function 
of the dipole tilting angle and the layer distance. Finally, we demonstrate in detail how the  
interlayer pairing gradually changes from being purely in the singlet channel when the dipoles 
are perpendicular to the layers, to being purely in the triplet channel when they are parallel to the layers. 

\section{System}
\begin{figure}
    \centering
    \begin{tikzpicture}[scale=.7]
    \draw[very thick] (0,0) -- (8,0);
    \draw[very thick] (0,2) -- (8,2);
    \foreach \x in {1,2,3,4,5,6,7} {%
      \draw[->, thick, >=latex] (\x,0) ++(-130:.5) -- ++(50:1);
      \draw[->, thick, >=latex] (\x,2) ++(-130:.5) -- ++(50:1);
    }
    \draw[|<->|, thin] (8.5,0) -- ++(0,2) node[pos=.5, right]{$\lambda$};
    \draw[|<->|, thin] (-.5,-.2) -- ++(0,.4) node[pos=.5, left]{$l_0$};
    \draw[|<->|, thin] (-.5,1.8) -- ++(0,.4) node[pos=.5, left]{$l_0$};
    \draw[->, thin, densely dotted] (3,-.5) -- ++(0,3.5) node[right]{$z$};
    \draw[thin, densely dotted] (3,0) -- ++(50:1.6);
    \draw[thin, densely dotted, ->] (3,0) ++(0,1.4) arc (90:50:1.4);
    \draw (3,0) ++(70:1) node{$\theta$};
    \draw (4,2) ++(50:.5) node[above]{$\mathbf{d}$};
    %\draw (8,1) node[right]{$\hat{\vec{d}} = \begin{pmatrix} \sin(\Theta) \\ 0 \\ \cos(\Theta) \end{pmatrix}$};
    \draw[blue!80!white, thick] (1.5,0) ellipse (1.1 and .7);
    \draw[blue!80!white, thick] (1.5,2) ellipse (1.1 and .7);
    \draw[blue!80!white, thick] (0,1) node[right]{$V_{\uparrow\uparrow}$};
    \draw[red!80!white, thick, shift={(5.5,1)}, rotate=60] (0,0) ellipse (1.8 and .3); 
    \draw[red!80!white, thick, shift={(6.5,1)}, rotate=60] (0,0) ellipse (1.8 and .3); 
    \draw[red!80!white] (7,1) node[right]{$V_{\uparrow\downarrow}$};
\end{tikzpicture}
    \caption{Dipoles are moving in two layers separated by the distance $\lambda$. An external field aligns the dipoles so that they 
    form an angle $\theta$ with respect to the normal of the planes. Intralayer/interlayer Cooper pairs driven by the interaction $V_{\uparrow\uparrow}$/$V_{\uparrow\downarrow}$
    are indicated by blue and red ellipses respectively. }
    \label{SetupFig}
\end{figure}
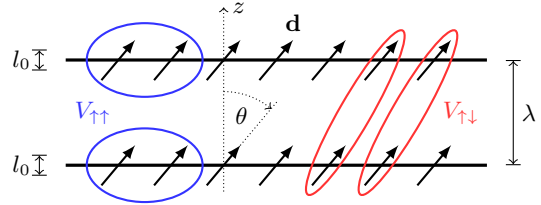

We consider dipolar fermions of mass $m$ which reside in two  parallel layers separated by a distance $\lambda$. 
The dipoles are confined in each layer by a strong harmonic potential perpendicular to the layer ($z$-direction), 
which ``freezes'' the dipoles in the lowest harmonic oscillator  state in the $z$-direction 
$\phi(z-z_s)=\exp[-(z-z_s)^2/2l_0^2]\pi^{-1/4}l_0^{-1/2}$, where $z_s$ is the position 
of layer $s$ along the $z$-direction and  $l_0$ is the layer thickness. This confinement makes the system effectively two-dimensional (2D).
The density of dipoles is $n=k_F^2/4\pi$ in each layer where $k_F$ is the Fermi momentum, and we denote dipoles in the upper layer as (pseudo) spin $\uparrow$ and dipoles in the 
lower layer as spin $\downarrow$. The dipoles are aligned by an external field so that their dipole moment ${\mathbf d}$ lies in the $xz$-plane 
forming an angle $\theta$ with the normal to the layers. We illustrate the setup in Fig.\ \ref{SetupFig}.
%\begin{figure}
%    \centering
%    \input{img/pairings.tex}
%    \caption{intra- and interlayer interaction \georg{Merge Fig. 1 and 2}}
%\end{figure}
The dipole-dipole interaction is $V({\mathbf r})  = D^2(1-3 \cos^2\theta_{rd})/r^3$, 
where  $\theta_{rd}$ is the angle between $\mathbf  d$ and 
the relative displacement vector ${\mathbf r}$ of the two dipoles. 
We have  $D^2 = d^2/4\pi\epsilon_0$ for electric dipoles and $D^2 = d^2\mu_0/4\pi$ for magnetic ones.

\subsection{Effective 2D Hamiltonian}
The effective interaction between  dipoles in  layer $s$ and $s'$ is obtained by integrating the dipole interaction $V({\mathbf r})$ over the 
 Gaussians  centered in each layer:  $V_{ss'}(\mathbf r_\perp-\mathbf r'_\perp)=\int\! dz \int\! dz'\phi(z-z_s) V(\mathbf{r}-\mathbf r')\phi(z'-z_{s'})$, where 
 $\mathbf r_\perp=(x,y)$ is the 2D position of a dipole in a layer.  Importantly, we see that 
  the interlayer interaction is   \emph{not} inversion symmetric, i.e.\ 
  $V_{\uparrow\downarrow}(-{\mathbf r}_\perp)\neq V_{\uparrow\downarrow}({\mathbf r}_\perp)$ for $0<\theta<\pi/2$. 
  Instead, it obeys the symmetry  
    $V_{\uparrow\downarrow}(-{\mathbf r}_\perp)= V_{\downarrow\uparrow}({\mathbf r}_\perp)$ where the two layers are interchanged, and 
    inversion symmetry is recovered only for $\theta=0$ and $\theta=\pi/2$. 
    On the other hand, the  intralayer interaction is always inversion symmetric with 
   $V_{\downarrow\downarrow}({\mathbf r}_\perp)=V_{\uparrow\uparrow}({\mathbf r}_\perp)= V_{\uparrow\uparrow}(-{\mathbf r}_\perp)$.
Performing a 2D Fourier transform yields~\cite{Fischer2006}
\begin{equation}
V_{\uparrow\uparrow}(\mathbf q)=-2\pi D^2F(l_0q)[\cos^2\theta-\sin^2\theta\cos^2\phi]
\label{eq:V2Dq}
\end{equation}
for the intralayer interaction, where  $F(x)= q\exp(x^2/2)\rm{erfc}(x/\sqrt{2})$ and $\mathbf q=(q_x,q_y)$. Here, $\phi$ is the azimuthal angle
of ${\mathbf q}$, and we have ignored a constant term (depending on $\theta$) in Eq.\  (\ref{eq:V2Dq}), since it plays no role for identical fermions.
The interlayer interaction is~\cite{Qiuzi2010} 
\begin{equation}
  \label{eq:Vinter}
 V_{\uparrow\downarrow}({\mathbf q})=2\pi D^2(i\cos\theta-\sin\theta\cos\phi)^2q e^{-\lambda q}
\end{equation}
for $l_0\ll \lambda$. We have  $V_{\uparrow\downarrow}(-{\mathbf k})= V_{\uparrow\downarrow}({\mathbf k})^*$.

These interactions give the  effective 2D Hamiltonian for the bi-layer system 
\begin{gather}
H=\sum_{\mathbf k,s}\frac{k^2}{2m}c^\dagger_{\mathbf ks}c_{\mathbf ks}
%\nonumber\\
+\frac{1}{{\mathcal V}}\sum_{\mathbf k,\mathbf k',\mathbf q}V_{\uparrow\downarrow}(\mathbf q)c^\dagger_{\mathbf k+\mathbf q\uparrow}
c^\dagger_{\mathbf k'-\mathbf q\downarrow}c_{\mathbf k'\downarrow}c_{\mathbf k\uparrow}\nonumber\\
+\frac{1}{2{\mathcal V}}\sum_{s}\sum_{\mathbf k,\mathbf k',\mathbf q}V_{\uparrow\uparrow}(\mathbf q)c^\dagger_{\mathbf k+\mathbf qs}
c^\dagger_{\mathbf k'-\mathbf qs}c_{\mathbf k's}c_{\mathbf ks},
  \label{Hamiltonian}
\end{gather}
where we have used $V_{\uparrow\downarrow}(-{\mathbf q})= V_{\downarrow\uparrow}({\mathbf q})$. Here, 
${\mathcal V}$ is the volume of the system,  and $ c_{\mathbf ks}$ removes a dipole in layer $s=\uparrow,\downarrow$ with 2D momentum $\mathbf k$.

\section{Pairing}
The  intra- and interlayer interaction has attractive regions, and we will now examine the resulting 
Cooper pairing between dipoles residing in the same layer (intralayer pairing) as 
well as pairing between dipoles residing in the two different layers (interlayer pairing).

\subsection{Interlayer pairing}
The interaction between dipoles in the two different layers has attractive regions for all dipole angles $\theta$. For the special case 
of the dipoles perpendicular to planes ($\theta=0$), the interlayer pairing has been examined using BCS theory without~\cite{Pikovski2010,Zinner2012}
 and with  induced interactions~\cite{Baranov2011}, with variational methods~\cite{Camacho2015}, as well as with Monte-Carlo methods~\cite{Matveeva2014}. 
 Here, we investigate the nature of the interlayer pairing for a general $\theta$. To do this, we use  BCS theory  
introducing the anomalous average 
$\langle c_{-{\mathbf k}\downarrow}c_{{\mathbf k}\uparrow}\rangle$, which describes interlayer pairing for a translationally invariant system. 
Defining  
$\Delta({\mathbf k})={\mathcal V}^{-1}\sum_{\mathbf k'}V_{\uparrow\downarrow}({\mathbf k}-{\mathbf k}')\langle a_{-{\mathbf k}'\downarrow}a_{{\mathbf k}'\uparrow}\rangle$, 
the  gap equation  becomes 
\begin{align}
\Delta({\mathbf k})=-\frac 1 {\mathcal V}\sum_{\mathbf k'}V_{\uparrow\downarrow}({\mathbf k}-{\mathbf k}')\frac{\Delta({\mathbf k}')}{2E_{\mathbf k'}},
\label{GapEq}
\end{align}
where $E_{\mathbf k}=\sqrt{\xi_k^2+|\Delta({\mathbf k})|^2}$ and $\xi_{\mathbf k}=k^2/2m-\mu$ with $\mu$ the chemical potential. 
Due to Fermi statistics, we can write the anomalous average as 
\begin{align}
\langle a_{-{\mathbf k}s'}a_{{\mathbf k}s}\rangle=A\phi_t({\mathbf k})\cdot\sigma_x+B\phi_s({\mathbf k})\cdot\sigma_y,
\label{Symmetry}
\end{align}
where  $\sigma_x=\left(\begin{smallmatrix*}   0 & 1 \\ 
    1& 0 
\end{smallmatrix*}\right)$ and  $\sigma_y=\left(\begin{smallmatrix*}   0 & -i \\ 
    i& 0 
\end{smallmatrix*}\right)$ are the Pauli matrices,  and $A$ and  $B$ are constants. 
The coefficient $A$ corresponds to interlayer triplet pairing with $\phi_t({\mathbf k})=-\phi_t(-{\mathbf k})$, and $B$ corresponds to interlayer 
singlet pairing with $\phi_s({\mathbf k})=\phi_s(-{\mathbf k})$. Since $V_{\uparrow\downarrow}({\mathbf k})\neq V_{\uparrow\downarrow}(-{\mathbf k})$
 for $0<\theta<\pi/2$, it follows from Eq.\ (\ref{GapEq}) that $\Delta({\mathbf k})\neq \pm\Delta(-{\mathbf k})$. Thus, a pure singlet or triplet pair wave function 
 is not a solution to Eq.\ (\ref{GapEq}), and the interlayer pairing  has both $A$ and $B$  different from zero corresponding to a mixed parity.    Note that this
  mixed parity pairing is due to an interaction without 
   inversion symmetry, in contrast to the case of non-centrosymmetric metals where the possibility of 
    mixed parity pairing arises due to a lack of inversion symmetry of the underlying crystal lattice~\cite{Sigrist2009}.

\subsection{Intralayer pairing}
The intra-layer interaction between two dipoles residing in the same layer has been shown to give rise to  pairing beyond 
a critical tilting angle $\theta_c\simeq\arcsin(2/3)\simeq0.23\pi$~\cite{Bruun2008}. 
We include this pairing instability by the  anomalous averages 
$\langle c_{-{\mathbf k}\uparrow}c_{{\mathbf k}\uparrow}\rangle$ and $\langle c_{-{\mathbf k}\downarrow}c_{{\mathbf k}\downarrow}\rangle$. 
The intralayer pairing is in the triplet channel, and  for a single layer, numerical calculations show that it is predominantly of $p$-wave 
character with a small $f$-wave component for some angles $\theta$~\cite{Bruun2008,Sieberer2011,Wu2015}. The gap equation for intralayer pairing is 
obtained from Eq.\ (\ref{GapEq}) simply by replacing $V_{\uparrow\downarrow}$ with $V_{\uparrow\uparrow}$.

\subsection{Eigenvalue analysis}
We will investigate  the competition between  inter- and intralayer pairing and in particular, what the ground state of the system is for different tilting 
angles $\theta$ and layer distances $\lambda$. To do this, we 
could solve in principle solve the corresponding non-linear gap equations taking a pairing field with a general symmetry. 
  There is fortunately a much simpler and more accurate approach. A considerable effort 
has in  recent years been put into the mathematical analysis \cite{HHSS, FHNS, HS,HaSe} of the gap-equation \eqref{GapEq}  
for rather  general pair interactions $V({\mathbf k}-{\mathbf k}')$,
in the thermodynamic limit ${\mathcal V} \to \infty$. For a comprehensive review see \cite{CompReview}.
Among other things it was rigorously proven that, in the weak coupling limit, the critical temperature \cite{FHNS,HS}, as well as the gap \cite{HS} $\Delta({\mathbf k_F})$
on the Fermi surface,
depends only on the lowest eigenvalue of an appropriate operator acting on functions which are living on the Fermi-surface. There is only pairing if
this eigenvalue is negative. For a circular Fermi-surface in 2D this operator has the specific form
\begin{equation} \label{eq:integral_operator}
%A_Vu(\varphi) = \frac{1}{4\pi} \int_{0}^{2\pi} V\left(k_F\tbinom{\cos\varphi-\cos\varphi'}{\sin\varphi-\sin\varphi'}\right)u(\varphi')\,\mathrm{d}\varphi' \, .
A_Vu(\varphi) = \frac{1}{4\pi} \int_{0}^{2\pi} V(\varphi,\varphi')u(\varphi')\,\mathrm{d}\varphi' \, ,
\end{equation}
where $V(\varphi,\varphi')\equiv V(k_F(\cos\varphi-\cos\varphi'),k_F(\sin\varphi-\sin\varphi'))$ is the interaction potential between momenta at the Fermi surface with 
azimuthal angles $\varphi$ and   $\varphi'$. 
The lowest eigenvalue $e(V)$ of the operator $A_V$ plays the role of the scattering length in the sense that \cite{FHNS,HS}
\begin{equation}
    \label{eq:Tc}
    T_c(V) \simeq \frac{8 e^{\gamma - 2}}{\pi}  \, e^{\frac{1}{k_F\, e(V)}}.
\end{equation}

Since the operator $A_V$ given by Eq.\ (\ref{eq:integral_operator}) is linear and acts on functions living on the Fermi surface only, it is 
 numerically much easier to analyse, compared to solving the full  non-linear gap equation in the whole of $\mathbf k$-space.
 Indeed,  our approach based on Eq.\ (\ref{eq:integral_operator}) allows us to obtain accurate results in the weak 
coupling limit for both intra- and interlayer pairing without too much numerical effort.  The numerical method 
  is described in App.\ \ref{Numerics}. 

\section{Phase diagram}
We now use this eigenvalue approach to examine the ground state of the system.
For all numerical calculations, we use the layer thickness $k_Fl_0=0.11$. 
Since the relative strength of the  intrashell and intershell pairing is independent of the dipole moment $D$, the nature of the 
ground state  depends only on  the two dimensionless parameters $\theta$ and $k_F\lambda$. For given values of $\theta$ and $k_F\lambda$,
the ground state is determined by the lowest negative eigenvalue of $A_V$: If  
$e(V_{\uparrow\downarrow})<e(V_{\uparrow\uparrow})$ the ground state is a superfluid with interlayer pairing, whereas the ground state 
is an intralayer superfluid if $e(V_{\uparrow\uparrow})<e(V_{\uparrow\downarrow})$.  Note that $e(V_{\uparrow\downarrow})<0$ for all $\theta$. 
The possibility of simultaneous interlayer and intralayer pairing is not considered here, since the presence of one order parameter  typically suppresses 
other kinds of order by gapping the Fermi surface.

\begin{figure}
    \centering
    % phase diagram for varying layer distance
% new without factor 1/2
\begin{tikzpicture}[>=latex, scale=1]
    % xscale=9.2000, xmin=0.0000, xmax=0.5000
    % yscale=3.1593, ymin=0.2240, ymax=1.6800
    % padding=0.2000, width=5.0000, height=5.0000
    \draw (0,0) rectangle (5.0000,5.0000);
    \draw (2.5000,0) node[below, scale=1.0]{$\theta/\pi$};
    \draw (0,2.5000) node[rotate=90, above, scale=1.0]{$k_F \lambda$};
    \foreach \x in {0.2000, 1.1200, 2.0400, 2.9600, 3.8800, 4.8000}{%
        \draw (\x, 0) -- ++(0, 0.1000);
        \draw (\x, 5.0000) -- ++(0, -0.1000);
    }
    \foreach \y in {0.2000, 2.5000, 4.8000}{%
        \draw (0, \y) -- ++(0.1000, 0);
        \draw (5.0000, \y) -- ++(-0.1000, 0);
    }
    \draw (0.2000, 0) node[below, scale=1.0]{$0.0$};
    \draw (4.8000, 0) node[below, scale=1.0]{$0.5$};
    \draw (0, 0.2000) node[left, scale=1.0]{$0.224$};
    \draw (0, 4.8000) node[left, scale=1.0]{$1.68$};
    \filldraw[thick, blue!20!white, shift={(0.2000, -0.5077)}] (0,.7077) rectangle (4.6,5.3077);
    \filldraw[thick, red!20!white] plot[shift={(0.2000, -0.5077)}] coordinates {%
        (0,.7077)
        (3.8400, 0.7077) (3.7794, 0.7744) (3.7234, 0.8410) (3.6711, 0.9077) (3.6220, 0.9744) (3.5756, 1.0410) (3.5315, 1.1077) (3.4896, 1.1744) (3.4495, 1.2410) (3.4111, 1.3077) (3.3743, 1.3744) (3.3388, 1.4410) (3.3047, 1.5077) (3.2718, 1.5744) (3.2400, 1.6410) (3.2093, 1.7077) (3.1796, 1.7744) (3.1509, 1.8410) (3.1230, 1.9077) (3.0960, 1.9744) (3.0698, 2.0410) (3.0444, 2.1077) (3.0197, 2.1744) (2.9958, 2.2410) (2.9725, 2.3077) (2.9499, 2.3744) (2.9279, 2.4410) (2.9065, 2.5077) (2.8857, 2.5744) (2.8654, 2.6410) (2.8458, 2.7077) (2.8266, 2.7744) (2.8080, 2.8410) (2.7899, 2.9077) (2.7722, 2.9744) (2.7550, 3.0410) (2.7383, 3.1077) (2.7220, 3.1744) (2.7062, 3.2410) (2.6907, 3.3077) (2.6757, 3.3744) (2.6611, 3.4410) (2.6468, 3.5077) (2.6330, 3.5744) (2.6194, 3.6410) (2.6063, 3.7077) (2.5935, 3.7744) (2.5810, 3.8410) (2.5688, 3.9077) (2.5570, 3.9744) (2.5455, 4.0410) (2.5343, 4.1077) (2.5233, 4.1744) (2.5127, 4.2410) (2.5023, 4.3077) (2.4922, 4.3744) (2.4824, 4.4410) (2.4728, 4.5077) (2.4635, 4.5744) (2.4544, 4.6410) (2.4456, 4.7077) (2.4369, 4.7744) (2.4285, 4.8410) (2.4204, 4.9077) (2.4124, 4.9744) (2.4046, 5.0410) (2.3971, 5.1077) (2.3897, 5.1744) (2.3825, 5.2410) (2.3756, 5.3077)
        (0, 5.3077) (0,.7077)
    };
    \draw[thick] plot[shift={(0.2000, -0.5077)}] coordinates {%
        (3.8400, 0.7077) (3.7794, 0.7744) (3.7234, 0.8410) (3.6711, 0.9077) (3.6220, 0.9744) (3.5756, 1.0410) (3.5315, 1.1077) (3.4896, 1.1744) (3.4495, 1.2410) (3.4111, 1.3077) (3.3743, 1.3744) (3.3388, 1.4410) (3.3047, 1.5077) (3.2718, 1.5744) (3.2400, 1.6410) (3.2093, 1.7077) (3.1796, 1.7744) (3.1509, 1.8410) (3.1230, 1.9077) (3.0960, 1.9744) (3.0698, 2.0410) (3.0444, 2.1077) (3.0197, 2.1744) (2.9958, 2.2410) (2.9725, 2.3077) (2.9499, 2.3744) (2.9279, 2.4410) (2.9065, 2.5077) (2.8857, 2.5744) (2.8654, 2.6410) (2.8458, 2.7077) (2.8266, 2.7744) (2.8080, 2.8410) (2.7899, 2.9077) (2.7722, 2.9744) (2.7550, 3.0410) (2.7383, 3.1077) (2.7220, 3.1744) (2.7062, 3.2410) (2.6907, 3.3077) (2.6757, 3.3744) (2.6611, 3.4410) (2.6468, 3.5077) (2.6330, 3.5744) (2.6194, 3.6410) (2.6063, 3.7077) (2.5935, 3.7744) (2.5810, 3.8410) (2.5688, 3.9077) (2.5570, 3.9744) (2.5455, 4.0410) (2.5343, 4.1077) (2.5233, 4.1744) (2.5127, 4.2410) (2.5023, 4.3077) (2.4922, 4.3744) (2.4824, 4.4410) (2.4728, 4.5077) (2.4635, 4.5744) (2.4544, 4.6410) (2.4456, 4.7077) (2.4369, 4.7744) (2.4285, 4.8410) (2.4204, 4.9077) (2.4124, 4.9744) (2.4046, 5.0410) (2.3971, 5.1077) (2.3897, 5.1744) (2.3825, 5.2410) (2.3756, 5.3077)
    };
    \draw[dashed] (2.27,0) -- (2.27,5);
    \draw (1.235,2.5) node[scale=1.0, text width=2cm, align=center]{interlayer pairing};
    \draw (3.890,2.5) node[scale=1.0, text width=2cm, align=center]{intralayer pairing};
    %\filldraw (2.922,2.652) circle(.075);
\end{tikzpicture}
    \caption{The ground state of the bi-layer system as a function of $\theta$ and $k_F\lambda$. For $\theta \leq \theta_c\simeq 0.23\pi$ (vertical dashed line), the interlayer pairing 
     always dominates since $A_{V_{\uparrow\uparrow}}$ has no negative eigenvalue. }
    \label{Phasediagram}
\end{figure}
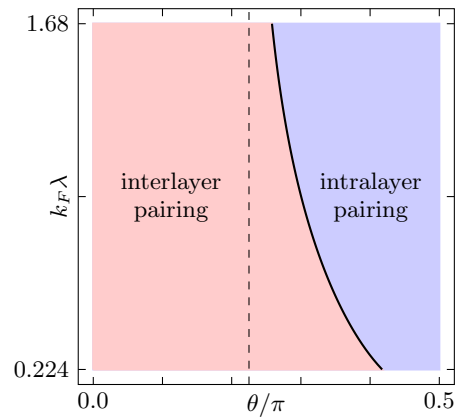
In Fig.\ \ref{Phasediagram}, we plot the resulting $T=0$ phase diagram.
For $\theta<\theta_c\simeq0.23\pi$, there is no intralayer pairing and the ground state is therefore a superfluid with interlayer pairing for all layer separations
 $k_F\lambda$. Of course, the critical temperature of the superfluid is vanishingly small for large  separations, since the interlayer interaction becomes 
very small. In the limit of large layer separation $k_F\lambda\gg1$, the intralayer pairing therefore wins as soon as $\theta>\theta_c$ as is 
indicated by the vertical dashed line in Fig.\ \ref{Phasediagram}. 
When the layer distance decreases and the strength of the interlayer interaction increases, 
Fig.\ \ref{Phasediagram} shows how  the interlayer superfluid is the ground state for increasingly large 
angles $\theta$ beyond $\theta_c$. We have not plotted the phase diagram for very small $k_F\lambda$ since the 
condition $l_0\ll\lambda$ for using Eq.\ (\ref{eq:Vinter}) breaks down in this regime.

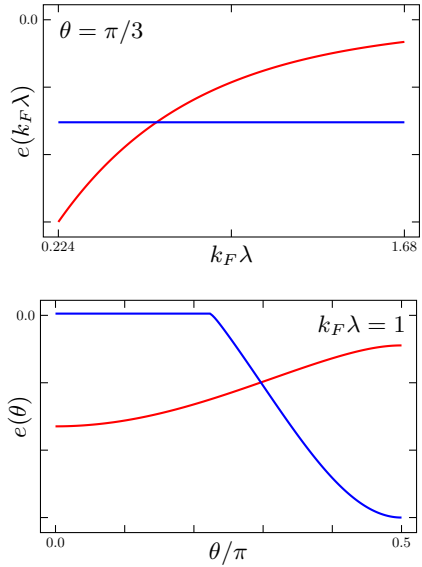
\begin{figure}
    \centering
    % e(kF lambda)
% new without factor 1/2
\begin{tikzpicture}[>=latex]
    \draw (0,0) rectangle (5.0000,3.0902);
    \draw (0.1,3.0) node[below right]{$\theta = \pi/3$};
    \draw (2.5000,0) node[below]{$k_F\lambda$};
    \draw (0,1.5451) node[rotate=90, above]{$e(k_F\lambda)$};
    \foreach \x in {0.2000, 2.5000, 4.8000}{%
        \draw (\x, 0) -- ++(0, 0.1000);
        \draw (\x, 3.0902) -- ++(0, -0.1000);
    }
    \foreach \y in {0.2000, 1.0967, 1.9935, 2.8902}{%
        \draw (0, \y) -- ++(0.1000, 0);
        \draw (5.0000, \y) -- ++(-0.1000, 0);
    }
    \draw (0.2000, 0) node[below, scale=.6]{$0.224$};
    \draw (4.8000, 0) node[below, scale=.6]{$1.68$};
    %\draw (0, 0.2000) node[left, scale=.6]{$-0.0022$};
    \draw (0, 2.8902) node[left, scale=.6]{$0.0$};
    \draw[thick, red] plot[smooth, shift={(-0.5077, 2.8902)}] coordinates {%
        (0.7077, -2.6902) (0.7384, -2.6466) (0.7690, -2.6038) (0.7997, -2.5617) (0.8304, -2.5203) (0.8610, -2.4796) (0.8917, -2.4397) (0.9224, -2.4004) (0.9530, -2.3618) (0.9837, -2.3238) (1.0144, -2.2865) (1.0450, -2.2499) (1.0757, -2.2138) (1.1064, -2.1784) (1.1370, -2.1436) (1.1677, -2.1093) (1.1984, -2.0757) (1.2290, -2.0426) (1.2597, -2.0101) (1.2904, -1.9782) (1.3210, -1.9467) (1.3517, -1.9158) (1.3824, -1.8855) (1.4130, -1.8556) (1.4437, -1.8263) (1.4744, -1.7975) (1.5050, -1.7691) (1.5357, -1.7412) (1.5664, -1.7138) (1.5970, -1.6869) (1.6277, -1.6604) (1.6584, -1.6343) (1.6890, -1.6087) (1.7197, -1.5836) (1.7504, -1.5588) (1.7810, -1.5345) (1.8117, -1.5106) (1.8424, -1.4870) (1.8730, -1.4639) (1.9037, -1.4412) (1.9344, -1.4188) (1.9650, -1.3968) (1.9957, -1.3752) (2.0264, -1.3540) (2.0570, -1.3331) (2.0877, -1.3125) (2.1184, -1.2923) (2.1490, -1.2724) (2.1797, -1.2529) (2.2104, -1.2337) (2.2410, -1.2148) (2.2717, -1.1962) (2.3024, -1.1780) (2.3330, -1.1600) (2.3637, -1.1423) (2.3944, -1.1250) (2.4250, -1.1079) (2.4557, -1.0911) (2.4864, -1.0745) (2.5170, -1.0583) (2.5477, -1.0423) (2.5784, -1.0266) (2.6090, -1.0111) (2.6397, -0.9959) (2.6704, -0.9810) (2.7010, -0.9663) (2.7317, -0.9518) (2.7624, -0.9376) (2.7930, -0.9236) (2.8237, -0.9098) (2.8544, -0.8963) (2.8850, -0.8829) (2.9157, -0.8698) (2.9464, -0.8570) (2.9770, -0.8443) (3.0077, -0.8318) (3.0384, -0.8196) (3.0690, -0.8075) (3.0997, -0.7956) (3.1304, -0.7840) (3.1610, -0.7725) (3.1917, -0.7612) (3.2224, -0.7501) (3.2530, -0.7391) (3.2837, -0.7284) (3.3144, -0.7178) (3.3450, -0.7074) (3.3757, -0.6971) (3.4064, -0.6871) (3.4370, -0.6771) (3.4677, -0.6674) (3.4984, -0.6578) (3.5290, -0.6484) (3.5597, -0.6391) (3.5904, -0.6299) (3.6210, -0.6209) (3.6517, -0.6121) (3.6824, -0.6034) (3.7130, -0.5948) (3.7437, -0.5864) (3.7744, -0.5781) (3.8050, -0.5699) (3.8357, -0.5619) (3.8664, -0.5540) (3.8970, -0.5462) (3.9277, -0.5386) (3.9584, -0.5310) (3.9890, -0.5236) (4.0197, -0.5163) (4.0504, -0.5091) (4.0810, -0.5021) (4.1117, -0.4951) (4.1424, -0.4883) (4.1730, -0.4815) (4.2037, -0.4749) (4.2344, -0.4684) (4.2650, -0.4620) (4.2957, -0.4556) (4.3264, -0.4494) (4.3570, -0.4433) (4.3877, -0.4373) (4.4184, -0.4313) (4.4490, -0.4255) (4.4797, -0.4197) (4.5104, -0.4141) (4.5410, -0.4085) (4.5717, -0.4030) (4.6024, -0.3976) (4.6330, -0.3923) (4.6637, -0.3871) (4.6944, -0.3819) (4.7250, -0.3769) (4.7557, -0.3719) (4.7864, -0.3670) (4.8170, -0.3621) (4.8477, -0.3574) (4.8784, -0.3527) (4.9090, -0.3481) (4.9397, -0.3435) (4.9704, -0.3390) (5.0010, -0.3346) (5.0317, -0.3303) (5.0624, -0.3260) (5.0930, -0.3218) (5.1237, -0.3177) (5.1544, -0.3136) (5.1850, -0.3096) (5.2157, -0.3056) (5.2464, -0.3017) (5.2770, -0.2979) (5.3077, -0.2941)
    };
    \draw[thick, blue] plot[smooth, shift={(-0.5077, 2.8902)}] coordinates {%
        (0.7077, -1.3639) (0.7384, -1.3639) (0.7690, -1.3639) (0.7997, -1.3639) (0.8304, -1.3639) (0.8610, -1.3639) (0.8917, -1.3639) (0.9224, -1.3639) (0.9530, -1.3639) (0.9837, -1.3639) (1.0144, -1.3639) (1.0450, -1.3639) (1.0757, -1.3639) (1.1064, -1.3639) (1.1370, -1.3639) (1.1677, -1.3639) (1.1984, -1.3639) (1.2290, -1.3639) (1.2597, -1.3639) (1.2904, -1.3639) (1.3210, -1.3639) (1.3517, -1.3639) (1.3824, -1.3639) (1.4130, -1.3639) (1.4437, -1.3639) (1.4744, -1.3639) (1.5050, -1.3639) (1.5357, -1.3639) (1.5664, -1.3639) (1.5970, -1.3639) (1.6277, -1.3639) (1.6584, -1.3639) (1.6890, -1.3639) (1.7197, -1.3639) (1.7504, -1.3639) (1.7810, -1.3639) (1.8117, -1.3639) (1.8424, -1.3639) (1.8730, -1.3639) (1.9037, -1.3639) (1.9344, -1.3639) (1.9650, -1.3639) (1.9957, -1.3639) (2.0264, -1.3639) (2.0570, -1.3639) (2.0877, -1.3639) (2.1184, -1.3639) (2.1490, -1.3639) (2.1797, -1.3639) (2.2104, -1.3639) (2.2410, -1.3639) (2.2717, -1.3639) (2.3024, -1.3639) (2.3330, -1.3639) (2.3637, -1.3639) (2.3944, -1.3639) (2.4250, -1.3639) (2.4557, -1.3639) (2.4864, -1.3639) (2.5170, -1.3639) (2.5477, -1.3639) (2.5784, -1.3639) (2.6090, -1.3639) (2.6397, -1.3639) (2.6704, -1.3639) (2.7010, -1.3639) (2.7317, -1.3639) (2.7624, -1.3639) (2.7930, -1.3639) (2.8237, -1.3639) (2.8544, -1.3639) (2.8850, -1.3639) (2.9157, -1.3639) (2.9464, -1.3639) (2.9770, -1.3639) (3.0077, -1.3639) (3.0384, -1.3639) (3.0690, -1.3639) (3.0997, -1.3639) (3.1304, -1.3639) (3.1610, -1.3639) (3.1917, -1.3639) (3.2224, -1.3639) (3.2530, -1.3639) (3.2837, -1.3639) (3.3144, -1.3639) (3.3450, -1.3639) (3.3757, -1.3639) (3.4064, -1.3639) (3.4370, -1.3639) (3.4677, -1.3639) (3.4984, -1.3639) (3.5290, -1.3639) (3.5597, -1.3639) (3.5904, -1.3639) (3.6210, -1.3639) (3.6517, -1.3639) (3.6824, -1.3639) (3.7130, -1.3639) (3.7437, -1.3639) (3.7744, -1.3639) (3.8050, -1.3639) (3.8357, -1.3639) (3.8664, -1.3639) (3.8970, -1.3639) (3.9277, -1.3639) (3.9584, -1.3639) (3.9890, -1.3639) (4.0197, -1.3639) (4.0504, -1.3639) (4.0810, -1.3639) (4.1117, -1.3639) (4.1424, -1.3639) (4.1730, -1.3639) (4.2037, -1.3639) (4.2344, -1.3639) (4.2650, -1.3639) (4.2957, -1.3639) (4.3264, -1.3639) (4.3570, -1.3639) (4.3877, -1.3639) (4.4184, -1.3639) (4.4490, -1.3639) (4.4797, -1.3639) (4.5104, -1.3639) (4.5410, -1.3639) (4.5717, -1.3639) (4.6024, -1.3639) (4.6330, -1.3639) (4.6637, -1.3639) (4.6944, -1.3639) (4.7250, -1.3639) (4.7557, -1.3639) (4.7864, -1.3639) (4.8170, -1.3639) (4.8477, -1.3639) (4.8784, -1.3639) (4.9090, -1.3639) (4.9397, -1.3639) (4.9704, -1.3639) (5.0010, -1.3639) (5.0317, -1.3639) (5.0624, -1.3639) (5.0930, -1.3639) (5.1237, -1.3639) (5.1544, -1.3639) (5.1850, -1.3639) (5.2157, -1.3639) (5.2464, -1.3639) (5.2770, -1.3639) (5.3077, -1.3639)
    };
\end{tikzpicture}\\[1em]
    % e(theta)
% new without factor 1/2
\begin{tikzpicture}[>=latex]
    \draw (0,0) rectangle (5.0000,3.0902);
    \draw (2.5000,0) node[below]{$\theta/\pi$};
    \draw (0,1.5451) node[rotate=90, above]{$e(\theta)$};
    \draw (4.97, 3.0) node[below left]{$k_F \lambda = 1$};
    \foreach \x in {0.2000, 1.1200, 2.0400, 2.9600, 3.8800, 4.8000}{%
        \draw (\x, 0) -- ++(0, 0.1000);
        \draw (\x, 3.0902) -- ++(0, -0.1000);
    }
    \foreach \y in {0.2000, 1.0967, 1.9935, 2.8902}{%
        \draw (0, \y) -- ++(0.1000, 0);
        \draw (5.0000, \y) -- ++(-0.1000, 0);
    }
    \draw (0.2000, 0) node[below, scale=.6]{$0.0$};
    \draw (4.8000, 0) node[below, scale=.6]{$0.5$};
    % \draw (0, 0.2000) node[left, scale=.6]{$-0.0022$};
    \draw (0, 2.8902) node[left, scale=.6]{$0.0$};
    \draw[thick, red] plot[smooth, shift={(0.2000, 2.9133)}] coordinates {%
        (0.0000, -1.4998) (0.0307, -1.4998) (0.0613, -1.4995) (0.0920, -1.4991) (0.1227, -1.4984) (0.1533, -1.4977) (0.1840, -1.4967) (0.2147, -1.4956) (0.2453, -1.4942) (0.2760, -1.4928) (0.3067, -1.4911) (0.3373, -1.4893) (0.3680, -1.4873) (0.3987, -1.4851) (0.4293, -1.4827) (0.4600, -1.4802) (0.4907, -1.4775) (0.5213, -1.4747) (0.5520, -1.4717) (0.5827, -1.4685) (0.6133, -1.4651) (0.6440, -1.4616) (0.6747, -1.4579) (0.7053, -1.4540) (0.7360, -1.4500) (0.7667, -1.4460) (0.7973, -1.4417) (0.8280, -1.4372) (0.8587, -1.4325) (0.8893, -1.4276) (0.9200, -1.4226) (0.9507, -1.4175) (0.9813, -1.4122) (1.0120, -1.4067) (1.0427, -1.4011) (1.0733, -1.3954) (1.1040, -1.3895) (1.1347, -1.3834) (1.1653, -1.3772) (1.1960, -1.3709) (1.2267, -1.3644) (1.2573, -1.3577) (1.2880, -1.3510) (1.3187, -1.3441) (1.3493, -1.3370) (1.3800, -1.3299) (1.4107, -1.3225) (1.4413, -1.3151) (1.4720, -1.3075) (1.5027, -1.2999) (1.5333, -1.2920) (1.5640, -1.2841) (1.5947, -1.2761) (1.6253, -1.2679) (1.6560, -1.2596) (1.6867, -1.2512) (1.7173, -1.2427) (1.7480, -1.2340) (1.7787, -1.2252) (1.8093, -1.2164) (1.8400, -1.2075) (1.8707, -1.1984) (1.9013, -1.1893) (1.9320, -1.1801) (1.9627, -1.1708) (1.9933, -1.1614) (2.0240, -1.1519) (2.0547, -1.1423) (2.0853, -1.1326) (2.1160, -1.1229) (2.1467, -1.1131) (2.1773, -1.1032) (2.2080, -1.0932) (2.2387, -1.0832) (2.2693, -1.0731) (2.3000, -1.0629) (2.3307, -1.0527) (2.3613, -1.0425) (2.3920, -1.0321) (2.4227, -1.0218) (2.4533, -1.0113) (2.4840, -1.0009) (2.5147, -0.9903) (2.5453, -0.9798) (2.5760, -0.9692) (2.6067, -0.9586) (2.6373, -0.9479) (2.6680, -0.9372) (2.6987, -0.9265) (2.7293, -0.9158) (2.7600, -0.9051) (2.7907, -0.8943) (2.8213, -0.8836) (2.8520, -0.8728) (2.8827, -0.8621) (2.9133, -0.8513) (2.9440, -0.8405) (2.9747, -0.8298) (3.0053, -0.8191) (3.0360, -0.8083) (3.0667, -0.7976) (3.0973, -0.7870) (3.1280, -0.7763) (3.1587, -0.7657) (3.1893, -0.7551) (3.2200, -0.7446) (3.2507, -0.7341) (3.2813, -0.7237) (3.3120, -0.7133) (3.3427, -0.7030) (3.3733, -0.6927) (3.4040, -0.6825) (3.4347, -0.6724) (3.4653, -0.6623) (3.4960, -0.6523) (3.5267, -0.6425) (3.5573, -0.6327) (3.5880, -0.6230) (3.6187, -0.6134) (3.6493, -0.6039) (3.6800, -0.5946) (3.7107, -0.5854) (3.7413, -0.5763) (3.7720, -0.5673) (3.8027, -0.5585) (3.8333, -0.5499) (3.8640, -0.5414) (3.8947, -0.5331) (3.9253, -0.5250) (3.9560, -0.5171) (3.9867, -0.5094) (4.0173, -0.5019) (4.0480, -0.4947) (4.0787, -0.4877) (4.1093, -0.4810) (4.1400, -0.4745) (4.1707, -0.4684) (4.2013, -0.4625) (4.2320, -0.4570) (4.2627, -0.4519) (4.2933, -0.4471) (4.3240, -0.4427) (4.3547, -0.4387) (4.3853, -0.4351) (4.4160, -0.4319) (4.4467, -0.4292) (4.4773, -0.4270) (4.5080, -0.4253) (4.5387, -0.4240) (4.5693, -0.4233) (4.6000, -0.4230)
    };
    \draw[thick, blue] plot[shift={(0.2000, 2.9133)}] coordinates {%
        (0.0000, 0.0000) (2.0547, 0.0000) (2.0853, -0.0241) (2.1160, -0.0554) (2.1467, -0.0916) (2.1773, -0.1302) (2.2080, -0.1703) (2.2387, -0.2113) (2.2693, -0.2530) (2.3000, -0.2952) (2.3307, -0.3378) (2.3613, -0.3808) (2.3920, -0.4242) (2.4227, -0.4679) (2.4533, -0.5117) (2.4840, -0.5558) (2.5147, -0.6001) (2.5453, -0.6446) (2.5760, -0.6892) (2.6067, -0.7340) (2.6373, -0.7788) (2.6680, -0.8237) (2.6987, -0.8687) (2.7293, -0.9137) (2.7600, -0.9587) (2.7907, -1.0037) (2.8213, -1.0486) (2.8520, -1.0935) (2.8827, -1.1383) (2.9133, -1.1830) (2.9440, -1.2275) (2.9747, -1.2719) (3.0053, -1.3160) (3.0360, -1.3600) (3.0667, -1.4037) (3.0973, -1.4472) (3.1280, -1.4903) (3.1587, -1.5331) (3.1893, -1.5756) (3.2200, -1.6177) (3.2507, -1.6596) (3.2813, -1.7009) (3.3120, -1.7417) (3.3427, -1.7821) (3.3733, -1.8219) (3.4040, -1.8612) (3.4347, -1.9000) (3.4653, -1.9381) (3.4960, -1.9756) (3.5267, -2.0125) (3.5573, -2.0487) (3.5880, -2.0842) (3.6187, -2.1190) (3.6493, -2.1530) (3.6800, -2.1863) (3.7107, -2.2187) (3.7413, -2.2503) (3.7720, -2.2811) (3.8027, -2.3109) (3.8333, -2.3399) (3.8640, -2.3679) (3.8947, -2.3950) (3.9253, -2.4211) (3.9560, -2.4462) (3.9867, -2.4703) (4.0173, -2.4934) (4.0480, -2.5154) (4.0787, -2.5363) (4.1093, -2.5561) (4.1400, -2.5748) (4.1707, -2.5924) (4.2013, -2.6089) (4.2320, -2.6241) (4.2627, -2.6382) (4.2933, -2.6512) (4.3240, -2.6629) (4.3547, -2.6734) (4.3853, -2.6827) (4.4160, -2.6908) (4.4467, -2.6977) (4.4773, -2.7033) (4.5080, -2.7077) (4.5387, -2.7108) (4.5693, -2.7127) (4.6000, -2.7133)
    };
\end{tikzpicture}
    \caption{Lowest eigenvalues $e(V_{\uparrow\downarrow})$ (red) and $e(V_{\uparrow\uparrow})$ (blue) as a function of the layer distance and the tilting angle respectively.}
    \label{fig:lowest_eigenvalues}
\end{figure}

To illustrate how the phase diagram is obtained, we plot in Fig.\ \ref{fig:lowest_eigenvalues} the lowest eigenvalues  $e(V_{\uparrow\downarrow})$ and 
$e(V_{\uparrow\uparrow})$  corresponding to interlayer and intralayer pairing respectively, as a 
function of $k_F\lambda$ for $\theta=\pi/3$, 
and  as a function of $\theta$ for $k_F\lambda=1$. The system exhibits a  quantum phase transition between inter- and intralayer pairing when 
the eigenvalues cross.  Note that it is crucial for obtaining $e(V_{\uparrow\uparrow})$ that only eigenvalues corresponding to  antisymmetric eigenfunctions are allowed.

\section{Symmetry of the interlayer pairing}
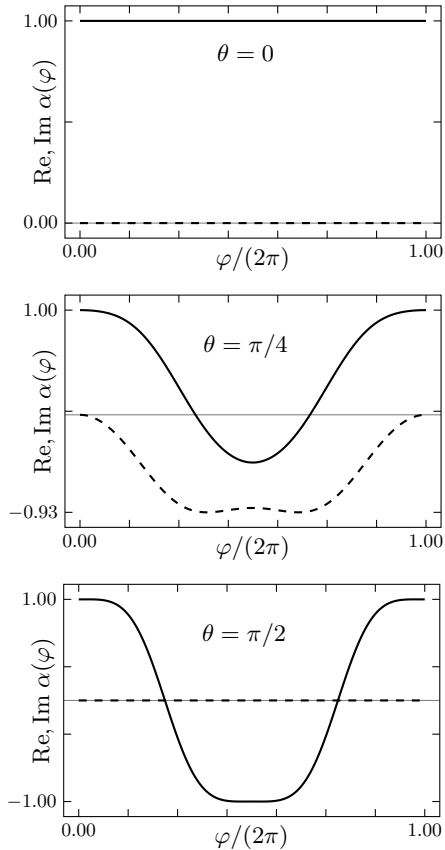
\begin{figure}
    \centering
    % re/im of (smallest) eigenfunction for th=0 without deformation
\begin{tikzpicture}[>=latex]
    \draw[gray] (0, 0.2000) -- ++(5.0000, 0);
    \draw (0,0) rectangle (5.0000,3.0902);
    \draw (2.5000,0) node[below]{$\varphi/(2\pi)$};
    \draw (0,1.5451) node[rotate=90, above]{$\mathrm{Re},\mathrm{Im} \; \alpha(\varphi)$};
    \foreach \x in {0.2000, 0.8571, 1.5143, 2.1714, 2.8286, 3.4857, 4.1429, 4.8000}{%
        \draw (\x, 0) -- ++(0, 0.1000);
        \draw (\x, 3.0902) -- ++(0, -0.1000);
    }
    \foreach \y in {0.2000, 1.5451, 2.8902}{%
        \draw (0, \y) -- ++(0.1000, 0);
        \draw (5.0000, \y) -- ++(-0.1000, 0);
    }
    \draw (0.2000, 0) node[below, scale=.8]{$0.00$};
    \draw (4.8000, 0) node[below, scale=.8]{$1.00$};
    \draw (0, 0.2000) node[left, scale=.8]{$\mathclap{\phantom{-}0.00\phantom{-0.00}}$};
    \draw (0, 2.8902) node[left, scale=.8]{$1.00$};
    \draw (2.4000, 2.6902) node[below, scale=1.0]{$\theta=0$};
    \draw[thick, black] plot[smooth, shift={(0.2000, 0.2000)}] coordinates {%
        (0.0000, 2.6902) (0.0307, 2.6902) (0.0613, 2.6902) (0.0920, 2.6902) (0.1227, 2.6902) (0.1533, 2.6902) (0.1840, 2.6902) (0.2147, 2.6902) (0.2453, 2.6902) (0.2760, 2.6902) (0.3067, 2.6902) (0.3373, 2.6902) (0.3680, 2.6902) (0.3987, 2.6902) (0.4293, 2.6902) (0.4600, 2.6902) (0.4907, 2.6902) (0.5213, 2.6902) (0.5520, 2.6902) (0.5827, 2.6902) (0.6133, 2.6902) (0.6440, 2.6902) (0.6747, 2.6902) (0.7053, 2.6902) (0.7360, 2.6902) (0.7667, 2.6902) (0.7973, 2.6902) (0.8280, 2.6902) (0.8587, 2.6902) (0.8893, 2.6902) (0.9200, 2.6902) (0.9507, 2.6902) (0.9813, 2.6902) (1.0120, 2.6902) (1.0427, 2.6902) (1.0733, 2.6902) (1.1040, 2.6902) (1.1347, 2.6902) (1.1653, 2.6902) (1.1960, 2.6902) (1.2267, 2.6902) (1.2573, 2.6902) (1.2880, 2.6902) (1.3187, 2.6902) (1.3493, 2.6902) (1.3800, 2.6902) (1.4107, 2.6902) (1.4413, 2.6902) (1.4720, 2.6902) (1.5027, 2.6902) (1.5333, 2.6902) (1.5640, 2.6902) (1.5947, 2.6902) (1.6253, 2.6902) (1.6560, 2.6902) (1.6867, 2.6902) (1.7173, 2.6902) (1.7480, 2.6902) (1.7787, 2.6902) (1.8093, 2.6902) (1.8400, 2.6902) (1.8707, 2.6902) (1.9013, 2.6902) (1.9320, 2.6902) (1.9627, 2.6902) (1.9933, 2.6902) (2.0240, 2.6902) (2.0547, 2.6902) (2.0853, 2.6902) (2.1160, 2.6902) (2.1467, 2.6902) (2.1773, 2.6902) (2.2080, 2.6902) (2.2387, 2.6902) (2.2693, 2.6902) (2.3000, 2.6902) (2.3307, 2.6902) (2.3613, 2.6902) (2.3920, 2.6902) (2.4227, 2.6902) (2.4533, 2.6902) (2.4840, 2.6902) (2.5147, 2.6902) (2.5453, 2.6902) (2.5760, 2.6902) (2.6067, 2.6902) (2.6373, 2.6902) (2.6680, 2.6902) (2.6987, 2.6902) (2.7293, 2.6902) (2.7600, 2.6902) (2.7907, 2.6902) (2.8213, 2.6902) (2.8520, 2.6902) (2.8827, 2.6902) (2.9133, 2.6902) (2.9440, 2.6902) (2.9747, 2.6902) (3.0053, 2.6902) (3.0360, 2.6902) (3.0667, 2.6902) (3.0973, 2.6902) (3.1280, 2.6902) (3.1587, 2.6902) (3.1893, 2.6902) (3.2200, 2.6902) (3.2507, 2.6902) (3.2813, 2.6902) (3.3120, 2.6902) (3.3427, 2.6902) (3.3733, 2.6902) (3.4040, 2.6902) (3.4347, 2.6902) (3.4653, 2.6902) (3.4960, 2.6902) (3.5267, 2.6902) (3.5573, 2.6902) (3.5880, 2.6902) (3.6187, 2.6902) (3.6493, 2.6902) (3.6800, 2.6902) (3.7107, 2.6902) (3.7413, 2.6902) (3.7720, 2.6902) (3.8027, 2.6902) (3.8333, 2.6902) (3.8640, 2.6902) (3.8947, 2.6902) (3.9253, 2.6902) (3.9560, 2.6902) (3.9867, 2.6902) (4.0173, 2.6902) (4.0480, 2.6902) (4.0787, 2.6902) (4.1093, 2.6902) (4.1400, 2.6902) (4.1707, 2.6902) (4.2013, 2.6902) (4.2320, 2.6902) (4.2627, 2.6902) (4.2933, 2.6902) (4.3240, 2.6902) (4.3547, 2.6902) (4.3853, 2.6902) (4.4160, 2.6902) (4.4467, 2.6902) (4.4773, 2.6902) (4.5080, 2.6902) (4.5387, 2.6902) (4.5693, 2.6902) (4.6000, 2.6902)
    };
    \draw[thick, black, dashed] plot[smooth, shift={(0.2000, 0.2000)}] coordinates {%
        (0.0000, -0.0000) (0.0307, -0.0000) (0.0613, -0.0000) (0.0920, -0.0000) (0.1227, -0.0000) (0.1533, -0.0000) (0.1840, -0.0000) (0.2147, -0.0000) (0.2453, -0.0000) (0.2760, -0.0000) (0.3067, -0.0000) (0.3373, -0.0000) (0.3680, -0.0000) (0.3987, -0.0000) (0.4293, -0.0000) (0.4600, -0.0000) (0.4907, -0.0000) (0.5213, -0.0000) (0.5520, -0.0000) (0.5827, -0.0000) (0.6133, -0.0000) (0.6440, -0.0000) (0.6747, -0.0000) (0.7053, -0.0000) (0.7360, -0.0000) (0.7667, -0.0000) (0.7973, -0.0000) (0.8280, -0.0000) (0.8587, -0.0000) (0.8893, -0.0000) (0.9200, -0.0000) (0.9507, -0.0000) (0.9813, -0.0000) (1.0120, -0.0000) (1.0427, -0.0000) (1.0733, -0.0000) (1.1040, -0.0000) (1.1347, -0.0000) (1.1653, -0.0000) (1.1960, -0.0000) (1.2267, -0.0000) (1.2573, -0.0000) (1.2880, -0.0000) (1.3187, -0.0000) (1.3493, -0.0000) (1.3800, -0.0000) (1.4107, -0.0000) (1.4413, -0.0000) (1.4720, -0.0000) (1.5027, -0.0000) (1.5333, -0.0000) (1.5640, -0.0000) (1.5947, -0.0000) (1.6253, -0.0000) (1.6560, -0.0000) (1.6867, -0.0000) (1.7173, -0.0000) (1.7480, -0.0000) (1.7787, -0.0000) (1.8093, -0.0000) (1.8400, -0.0000) (1.8707, -0.0000) (1.9013, -0.0000) (1.9320, -0.0000) (1.9627, -0.0000) (1.9933, -0.0000) (2.0240, -0.0000) (2.0547, -0.0000) (2.0853, -0.0000) (2.1160, -0.0000) (2.1467, -0.0000) (2.1773, -0.0000) (2.2080, -0.0000) (2.2387, -0.0000) (2.2693, -0.0000) (2.3000, -0.0000) (2.3307, -0.0000) (2.3613, -0.0000) (2.3920, -0.0000) (2.4227, -0.0000) (2.4533, -0.0000) (2.4840, -0.0000) (2.5147, -0.0000) (2.5453, -0.0000) (2.5760, -0.0000) (2.6067, -0.0000) (2.6373, -0.0000) (2.6680, -0.0000) (2.6987, -0.0000) (2.7293, -0.0000) (2.7600, -0.0000) (2.7907, -0.0000) (2.8213, -0.0000) (2.8520, -0.0000) (2.8827, -0.0000) (2.9133, -0.0000) (2.9440, -0.0000) (2.9747, -0.0000) (3.0053, -0.0000) (3.0360, -0.0000) (3.0667, -0.0000) (3.0973, -0.0000) (3.1280, -0.0000) (3.1587, -0.0000) (3.1893, -0.0000) (3.2200, -0.0000) (3.2507, -0.0000) (3.2813, -0.0000) (3.3120, -0.0000) (3.3427, -0.0000) (3.3733, -0.0000) (3.4040, -0.0000) (3.4347, -0.0000) (3.4653, -0.0000) (3.4960, -0.0000) (3.5267, -0.0000) (3.5573, -0.0000) (3.5880, -0.0000) (3.6187, -0.0000) (3.6493, -0.0000) (3.6800, -0.0000) (3.7107, -0.0000) (3.7413, -0.0000) (3.7720, -0.0000) (3.8027, -0.0000) (3.8333, -0.0000) (3.8640, -0.0000) (3.8947, -0.0000) (3.9253, -0.0000) (3.9560, -0.0000) (3.9867, -0.0000) (4.0173, -0.0000) (4.0480, -0.0000) (4.0787, -0.0000) (4.1093, -0.0000) (4.1400, -0.0000) (4.1707, -0.0000) (4.2013, -0.0000) (4.2320, -0.0000) (4.2627, -0.0000) (4.2933, -0.0000) (4.3240, -0.0000) (4.3547, -0.0000) (4.3853, -0.0000) (4.4160, -0.0000) (4.4467, -0.0000) (4.4773, -0.0000) (4.5080, -0.0000) (4.5387, -0.0000) (4.5693, -0.0000) (4.6000, 0.0000)
    };
\end{tikzpicture} \\[5pt]
    % re/im of (smallest) eigenfunction for th=0.785398 without deformation
\begin{tikzpicture}[>=latex]
    \draw[gray] (0, 1.4975) -- ++(5.0000, 0);
    \draw (0,0) rectangle (5.0000,3.0902);
    \draw (2.5000,0) node[below]{$\varphi/(2\pi)$};
    \draw (0,1.5451) node[rotate=90, above]{$\mathrm{Re},\mathrm{Im} \; \alpha(\varphi)$};
    \foreach \x in {0.2000, 0.8571, 1.5143, 2.1714, 2.8286, 3.4857, 4.1429, 4.8000}{%
        \draw (\x, 0) -- ++(0, 0.1000);
        \draw (\x, 3.0902) -- ++(0, -0.1000);
    }
    \foreach \y in {0.2000, 1.5451, 2.8902}{%
        \draw (0, \y) -- ++(0.1000, 0);
        \draw (5.0000, \y) -- ++(-0.1000, 0);
    }
    \draw (0.2000, 0) node[below, scale=.8]{$0.00$};
    \draw (4.8000, 0) node[below, scale=.8]{$1.00$};
    \draw (0, 0.2000) node[left, scale=.8]{$\mathclap{-0.93\phantom{-0.00}}$};
    \draw (0, 2.8902) node[left, scale=.8]{$1.00$};
    \draw (2.4000, 2.6902) node[below, scale=1.0]{$\theta=\pi/4$};
    \draw[thick, black] plot[smooth, shift={(0.2000, 1.4975)}] coordinates {%
        (0.0000, 1.3927) (0.0307, 1.3924) (0.0613, 1.3917) (0.0920, 1.3906) (0.1227, 1.3890) (0.1533, 1.3869) (0.1840, 1.3843) (0.2147, 1.3810) (0.2453, 1.3771) (0.2760, 1.3725) (0.3067, 1.3671) (0.3373, 1.3608) (0.3680, 1.3536) (0.3987, 1.3453) (0.4293, 1.3358) (0.4600, 1.3251) (0.4907, 1.3131) (0.5213, 1.2997) (0.5520, 1.2848) (0.5827, 1.2683) (0.6133, 1.2502) (0.6440, 1.2303) (0.6747, 1.2086) (0.7053, 1.1850) (0.7360, 1.1596) (0.7667, 1.1322) (0.7973, 1.1029) (0.8280, 1.0716) (0.8587, 1.0384) (0.8893, 1.0032) (0.9200, 0.9662) (0.9507, 0.9273) (0.9813, 0.8867) (1.0120, 0.8444) (1.0427, 0.8005) (1.0733, 0.7551) (1.1040, 0.7084) (1.1347, 0.6605) (1.1653, 0.6116) (1.1960, 0.5617) (1.2267, 0.5111) (1.2573, 0.4599) (1.2880, 0.4084) (1.3187, 0.3566) (1.3493, 0.3048) (1.3800, 0.2531) (1.4107, 0.2018) (1.4413, 0.1509) (1.4720, 0.1007) (1.5027, 0.0514) (1.5333, 0.0030) (1.5640, -0.0443) (1.5947, -0.0903) (1.6253, -0.1350) (1.6560, -0.1781) (1.6867, -0.2196) (1.7173, -0.2595) (1.7480, -0.2976) (1.7787, -0.3338) (1.8093, -0.3681) (1.8400, -0.4005) (1.8707, -0.4308) (1.9013, -0.4592) (1.9320, -0.4854) (1.9627, -0.5095) (1.9933, -0.5316) (2.0240, -0.5515) (2.0547, -0.5693) (2.0853, -0.5850) (2.1160, -0.5985) (2.1467, -0.6099) (2.1773, -0.6192) (2.2080, -0.6264) (2.2387, -0.6314) (2.2693, -0.6343) (2.3000, -0.6351) (2.3307, -0.6338) (2.3613, -0.6304) (2.3920, -0.6248) (2.4227, -0.6171) (2.4533, -0.6073) (2.4840, -0.5953) (2.5147, -0.5813) (2.5453, -0.5651) (2.5760, -0.5467) (2.6067, -0.5263) (2.6373, -0.5037) (2.6680, -0.4790) (2.6987, -0.4523) (2.7293, -0.4234) (2.7600, -0.3926) (2.7907, -0.3597) (2.8213, -0.3249) (2.8520, -0.2882) (2.8827, -0.2497) (2.9133, -0.2094) (2.9440, -0.1675) (2.9747, -0.1239) (3.0053, -0.0789) (3.0360, -0.0326) (3.0667, 0.0150) (3.0973, 0.0636) (3.1280, 0.1132) (3.1587, 0.1636) (3.1893, 0.2146) (3.2200, 0.2660) (3.2507, 0.3177) (3.2813, 0.3695) (3.3120, 0.4213) (3.3427, 0.4728) (3.3733, 0.5238) (3.4040, 0.5742) (3.4347, 0.6239) (3.4653, 0.6726) (3.4960, 0.7202) (3.5267, 0.7666) (3.5573, 0.8116) (3.5880, 0.8551) (3.6187, 0.8970) (3.6493, 0.9372) (3.6800, 0.9756) (3.7107, 1.0122) (3.7413, 1.0469) (3.7720, 1.0796) (3.8027, 1.1104) (3.8333, 1.1393) (3.8640, 1.1661) (3.8947, 1.1911) (3.9253, 1.2142) (3.9560, 1.2354) (3.9867, 1.2549) (4.0173, 1.2726) (4.0480, 1.2887) (4.0787, 1.3032) (4.1093, 1.3163) (4.1400, 1.3279) (4.1707, 1.3383) (4.2013, 1.3474) (4.2320, 1.3555) (4.2627, 1.3625) (4.2933, 1.3685) (4.3240, 1.3737) (4.3547, 1.3782) (4.3853, 1.3819) (4.4160, 1.3850) (4.4467, 1.3875) (4.4773, 1.3895) (4.5080, 1.3909) (4.5387, 1.3920) (4.5693, 1.3925) (4.6000, 1.3927)
    };
    \draw[thick, black, dashed] plot[smooth, shift={(0.2000, 1.4975)}] coordinates {%
        (0.0000, -0.0000) (0.0307, -0.0014) (0.0613, -0.0049) (0.0920, -0.0106) (0.1227, -0.0184) (0.1533, -0.0283) (0.1840, -0.0404) (0.2147, -0.0546) (0.2453, -0.0708) (0.2760, -0.0890) (0.3067, -0.1092) (0.3373, -0.1313) (0.3680, -0.1554) (0.3987, -0.1812) (0.4293, -0.2087) (0.4600, -0.2379) (0.4907, -0.2687) (0.5213, -0.3009) (0.5520, -0.3345) (0.5827, -0.3694) (0.6133, -0.4053) (0.6440, -0.4423) (0.6747, -0.4801) (0.7053, -0.5186) (0.7360, -0.5576) (0.7667, -0.5970) (0.7973, -0.6367) (0.8280, -0.6764) (0.8587, -0.7159) (0.8893, -0.7552) (0.9200, -0.7940) (0.9507, -0.8322) (0.9813, -0.8696) (1.0120, -0.9061) (1.0427, -0.9414) (1.0733, -0.9755) (1.1040, -1.0083) (1.1347, -1.0395) (1.1653, -1.0692) (1.1960, -1.0971) (1.2267, -1.1233) (1.2573, -1.1477) (1.2880, -1.1701) (1.3187, -1.1907) (1.3493, -1.2093) (1.3800, -1.2260) (1.4107, -1.2408) (1.4413, -1.2537) (1.4720, -1.2647) (1.5027, -1.2740) (1.5333, -1.2816) (1.5640, -1.2876) (1.5947, -1.2921) (1.6253, -1.2952) (1.6560, -1.2969) (1.6867, -1.2975) (1.7173, -1.2970) (1.7480, -1.2956) (1.7787, -1.2933) (1.8093, -1.2904) (1.8400, -1.2869) (1.8707, -1.2830) (1.9013, -1.2788) (1.9320, -1.2743) (1.9627, -1.2698) (1.9933, -1.2654) (2.0240, -1.2610) (2.0547, -1.2569) (2.0853, -1.2530) (2.1160, -1.2495) (2.1467, -1.2465) (2.1773, -1.2439) (2.2080, -1.2419) (2.2387, -1.2405) (2.2693, -1.2397) (2.3000, -1.2394) (2.3307, -1.2398) (2.3613, -1.2408) (2.3920, -1.2424) (2.4227, -1.2445) (2.4533, -1.2472) (2.4840, -1.2504) (2.5147, -1.2539) (2.5453, -1.2579) (2.5760, -1.2621) (2.6067, -1.2665) (2.6373, -1.2710) (2.6680, -1.2755) (2.6987, -1.2798) (2.7293, -1.2840) (2.7600, -1.2878) (2.7907, -1.2912) (2.8213, -1.2939) (2.8520, -1.2960) (2.8827, -1.2972) (2.9133, -1.2975) (2.9440, -1.2966) (2.9747, -1.2945) (3.0053, -1.2911) (3.0360, -1.2863) (3.0667, -1.2799) (3.0973, -1.2719) (3.1280, -1.2621) (3.1587, -1.2506) (3.1893, -1.2373) (3.2200, -1.2220) (3.2507, -1.2048) (3.2813, -1.1857) (3.3120, -1.1647) (3.3427, -1.1417) (3.3733, -1.1169) (3.4040, -1.0903) (3.4347, -1.0619) (3.4653, -1.0319) (3.4960, -1.0002) (3.5267, -0.9671) (3.5573, -0.9327) (3.5880, -0.8971) (3.6187, -0.8604) (3.6493, -0.8227) (3.6800, -0.7844) (3.7107, -0.7454) (3.7413, -0.7061) (3.7720, -0.6664) (3.8027, -0.6267) (3.8333, -0.5871) (3.8640, -0.5478) (3.8947, -0.5089) (3.9253, -0.4706) (3.9560, -0.4330) (3.9867, -0.3962) (4.0173, -0.3606) (4.0480, -0.3260) (4.0787, -0.2927) (4.1093, -0.2609) (4.1400, -0.2305) (4.1707, -0.2017) (4.2013, -0.1746) (4.2320, -0.1492) (4.2627, -0.1256) (4.2933, -0.1040) (4.3240, -0.0843) (4.3547, -0.0665) (4.3853, -0.0508) (4.4160, -0.0372) (4.4467, -0.0256) (4.4773, -0.0162) (4.5080, -0.0089) (4.5387, -0.0038) (4.5693, -0.0008) (4.6000, 0.0000)
    };
\end{tikzpicture} \\[5pt]
    \hspace*{2pt}% re/im of (smallest) eigenfunction for th=1.5708 without deformation
\begin{tikzpicture}[>=latex]
    \draw[gray] (0, 1.5451) -- ++(5.0000, 0);
    \draw (0,0) rectangle (5.0000,3.0902);
    \draw (2.5000,0) node[below]{$\varphi/(2\pi)$};
    \draw (0,1.5451) node[rotate=90, above]{$\mathrm{Re},\mathrm{Im} \; \alpha(\varphi)$};
    \foreach \x in {0.2000, 0.8571, 1.5143, 2.1714, 2.8286, 3.4857, 4.1429, 4.8000}{%
        \draw (\x, 0) -- ++(0, 0.1000);
        \draw (\x, 3.0902) -- ++(0, -0.1000);
    }
    \foreach \y in {0.2000, 1.0967, 1.9935, 2.8902}{%
        \draw (0, \y) -- ++(0.1000, 0);
        \draw (5.0000, \y) -- ++(-0.1000, 0);
    }
    \draw (0.2000, 0) node[below, scale=.8]{$0.00$};
    \draw (4.8000, 0) node[below, scale=.8]{$1.00$};
    \draw (0, 0.2000) node[left, scale=.8]{$\mathclap{-1.00\phantom{-0.00}}$};
    \draw (0, 2.8902) node[left, scale=.8]{$1.00$};
    \draw (2.4000, 2.6902) node[below, scale=1.0]{$\theta=\pi/2$};
    \draw[thick, black] plot[smooth, shift={(0.2000, 1.5451)}] coordinates {%
        (0.0000, 1.3448) (0.0307, 1.3449) (0.0613, 1.3450) (0.0920, 1.3451) (0.1227, 1.3451) (0.1533, 1.3450) (0.1840, 1.3446) (0.2147, 1.3438) (0.2453, 1.3424) (0.2760, 1.3402) (0.3067, 1.3371) (0.3373, 1.3327) (0.3680, 1.3269) (0.3987, 1.3193) (0.4293, 1.3096) (0.4600, 1.2976) (0.4907, 1.2829) (0.5213, 1.2651) (0.5520, 1.2441) (0.5827, 1.2194) (0.6133, 1.1908) (0.6440, 1.1579) (0.6747, 1.1207) (0.7053, 1.0788) (0.7360, 1.0321) (0.7667, 0.9805) (0.7973, 0.9240) (0.8280, 0.8625) (0.8587, 0.7963) (0.8893, 0.7254) (0.9200, 0.6502) (0.9507, 0.5708) (0.9813, 0.4878) (1.0120, 0.4015) (1.0427, 0.3125) (1.0733, 0.2214) (1.1040, 0.1288) (1.1347, 0.0352) (1.1653, -0.0586) (1.1960, -0.1520) (1.2267, -0.2444) (1.2573, -0.3350) (1.2880, -0.4234) (1.3187, -0.5089) (1.3493, -0.5910) (1.3800, -0.6694) (1.4107, -0.7436) (1.4413, -0.8133) (1.4720, -0.8784) (1.5027, -0.9386) (1.5333, -0.9939) (1.5640, -1.0442) (1.5947, -1.0897) (1.6253, -1.1304) (1.6560, -1.1666) (1.6867, -1.1983) (1.7173, -1.2259) (1.7480, -1.2497) (1.7787, -1.2699) (1.8093, -1.2868) (1.8400, -1.3008) (1.8707, -1.3122) (1.9013, -1.3214) (1.9320, -1.3285) (1.9627, -1.3340) (1.9933, -1.3380) (2.0240, -1.3409) (2.0547, -1.3428) (2.0853, -1.3440) (2.1160, -1.3447) (2.1467, -1.3450) (2.1773, -1.3451) (2.2080, -1.3450) (2.2387, -1.3449) (2.2693, -1.3449) (2.3000, -1.3448) (2.3307, -1.3449) (2.3613, -1.3450) (2.3920, -1.3451) (2.4227, -1.3451) (2.4533, -1.3450) (2.4840, -1.3446) (2.5147, -1.3438) (2.5453, -1.3424) (2.5760, -1.3402) (2.6067, -1.3371) (2.6373, -1.3327) (2.6680, -1.3269) (2.6987, -1.3193) (2.7293, -1.3096) (2.7600, -1.2976) (2.7907, -1.2829) (2.8213, -1.2651) (2.8520, -1.2441) (2.8827, -1.2194) (2.9133, -1.1908) (2.9440, -1.1579) (2.9747, -1.1207) (3.0053, -1.0788) (3.0360, -1.0321) (3.0667, -0.9805) (3.0973, -0.9240) (3.1280, -0.8625) (3.1587, -0.7963) (3.1893, -0.7254) (3.2200, -0.6502) (3.2507, -0.5708) (3.2813, -0.4878) (3.3120, -0.4015) (3.3427, -0.3125) (3.3733, -0.2214) (3.4040, -0.1288) (3.4347, -0.0352) (3.4653, 0.0586) (3.4960, 0.1520) (3.5267, 0.2444) (3.5573, 0.3350) (3.5880, 0.4234) (3.6187, 0.5089) (3.6493, 0.5910) (3.6800, 0.6694) (3.7107, 0.7436) (3.7413, 0.8133) (3.7720, 0.8784) (3.8027, 0.9386) (3.8333, 0.9939) (3.8640, 1.0442) (3.8947, 1.0897) (3.9253, 1.1304) (3.9560, 1.1666) (3.9867, 1.1983) (4.0173, 1.2259) (4.0480, 1.2497) (4.0787, 1.2699) (4.1093, 1.2868) (4.1400, 1.3008) (4.1707, 1.3122) (4.2013, 1.3214) (4.2320, 1.3285) (4.2627, 1.3340) (4.2933, 1.3380) (4.3240, 1.3409) (4.3547, 1.3428) (4.3853, 1.3440) (4.4160, 1.3447) (4.4467, 1.3450) (4.4773, 1.3451) (4.5080, 1.3450) (4.5387, 1.3449) (4.5693, 1.3449) (4.6000, 1.3448)
    };
    \draw[thick, black, dashed] plot[smooth, shift={(0.2000, 1.5451)}] coordinates {%
        (0.0000, 0.0000) (0.0307, -0.0000) (0.0613, -0.0000) (0.0920, -0.0000) (0.1227, -0.0000) (0.1533, -0.0000) (0.1840, -0.0000) (0.2147, -0.0000) (0.2453, -0.0000) (0.2760, -0.0000) (0.3067, -0.0000) (0.3373, -0.0000) (0.3680, -0.0000) (0.3987, -0.0000) (0.4293, -0.0000) (0.4600, -0.0000) (0.4907, -0.0000) (0.5213, -0.0000) (0.5520, -0.0000) (0.5827, -0.0000) (0.6133, -0.0000) (0.6440, -0.0000) (0.6747, -0.0000) (0.7053, -0.0000) (0.7360, -0.0000) (0.7667, -0.0000) (0.7973, -0.0000) (0.8280, -0.0000) (0.8587, -0.0000) (0.8893, -0.0000) (0.9200, -0.0000) (0.9507, -0.0000) (0.9813, -0.0000) (1.0120, -0.0000) (1.0427, -0.0000) (1.0733, -0.0000) (1.1040, -0.0000) (1.1347, -0.0000) (1.1653, -0.0000) (1.1960, -0.0000) (1.2267, -0.0000) (1.2573, -0.0000) (1.2880, -0.0000) (1.3187, -0.0000) (1.3493, -0.0000) (1.3800, -0.0000) (1.4107, -0.0000) (1.4413, -0.0000) (1.4720, -0.0000) (1.5027, -0.0000) (1.5333, -0.0000) (1.5640, -0.0000) (1.5947, -0.0000) (1.6253, -0.0000) (1.6560, -0.0000) (1.6867, -0.0000) (1.7173, -0.0000) (1.7480, -0.0000) (1.7787, 0.0000) (1.8093, 0.0000) (1.8400, 0.0000) (1.8707, 0.0000) (1.9013, 0.0000) (1.9320, 0.0000) (1.9627, 0.0000) (1.9933, 0.0000) (2.0240, 0.0000) (2.0547, 0.0000) (2.0853, 0.0000) (2.1160, 0.0000) (2.1467, 0.0000) (2.1773, 0.0000) (2.2080, 0.0000) (2.2387, 0.0000) (2.2693, 0.0000) (2.3000, 0.0000) (2.3307, 0.0000) (2.3613, 0.0000) (2.3920, 0.0000) (2.4227, 0.0000) (2.4533, 0.0000) (2.4840, 0.0000) (2.5147, 0.0000) (2.5453, 0.0000) (2.5760, 0.0000) (2.6067, 0.0000) (2.6373, 0.0000) (2.6680, 0.0000) (2.6987, 0.0000) (2.7293, 0.0000) (2.7600, 0.0000) (2.7907, 0.0000) (2.8213, 0.0000) (2.8520, -0.0000) (2.8827, -0.0000) (2.9133, -0.0000) (2.9440, -0.0000) (2.9747, -0.0000) (3.0053, -0.0000) (3.0360, -0.0000) (3.0667, -0.0000) (3.0973, -0.0000) (3.1280, -0.0000) (3.1587, -0.0000) (3.1893, -0.0000) (3.2200, -0.0000) (3.2507, -0.0000) (3.2813, -0.0000) (3.3120, -0.0000) (3.3427, -0.0000) (3.3733, -0.0000) (3.4040, -0.0000) (3.4347, -0.0000) (3.4653, -0.0000) (3.4960, -0.0000) (3.5267, -0.0000) (3.5573, -0.0000) (3.5880, -0.0000) (3.6187, -0.0000) (3.6493, -0.0000) (3.6800, -0.0000) (3.7107, -0.0000) (3.7413, -0.0000) (3.7720, -0.0000) (3.8027, -0.0000) (3.8333, -0.0000) (3.8640, -0.0000) (3.8947, -0.0000) (3.9253, -0.0000) (3.9560, -0.0000) (3.9867, -0.0000) (4.0173, -0.0000) (4.0480, -0.0000) (4.0787, -0.0000) (4.1093, -0.0000) (4.1400, -0.0000) (4.1707, -0.0000) (4.2013, -0.0000) (4.2320, -0.0000) (4.2627, -0.0000) (4.2933, -0.0000) (4.3240, -0.0000) (4.3547, -0.0000) (4.3853, -0.0000) (4.4160, -0.0000) (4.4467, -0.0000) (4.4773, -0.0000) (4.5080, -0.0000) (4.5387, -0.0000) (4.5693, -0.0000) (4.6000, 0.0000)
    };
\end{tikzpicture}
    \caption{Angular dependence of the eigenfunctions $\alpha$ of $A_{V_{\uparrow\downarrow}}$ (real parts solid, imaginary parts dashed) for $k_F\lambda =1$. 
    Since the dipole orientation is in the $xz$-plane, we always have $\alpha(\varphi)=\alpha(-\varphi)$.}
    \label{alphaSymmetry}
\end{figure}

We now discuss in more detail the symmetry of the  interlayer pairing. Figure \ref{alphaSymmetry} shows the interlayer 
pair wave function $\alpha(\varphi)$, which is the eigenfunction of $A_{V_{\uparrow\downarrow}}$ with the lowest eigenvalue, 
  as a function of the  azimuthal angle $\varphi$ on the Fermi surface, for three 
different tilting angles: $\theta=0$, $\theta=\pi/4$, and $\theta=\pi/2$. The layer distance is $k_F\lambda =1$. 
As expected, the pair wave function is purely singlet $s$-wave for $\theta=0$ reflecting that the interlayer interaction is rotationally symmetric 
when the dipoles are perpendicular to the layers. When $\theta=\pi/4$ on the other hand, we see that  the pair wave function 
has no definite parity with $\alpha(\varphi+\pi)\neq\pm\alpha(\varphi)$ corresponding to both  $A\neq0$ and $B\neq0$ in Eq.\ (\ref{Symmetry}).
The pairing is purely in the triplet channel with $\alpha(\varphi+\pi)=-\alpha(\varphi)$ for $\theta=\pi/2$, where the interlayer  
is inversion symmetric.   

To quantify the symmetry of the pair wave function further,  we split it into a symmetric and an anti-symmetric part writing 
\begin{equation}%\small
    \alpha(\varphi) = \underbrace{\frac{\alpha(\varphi) + \alpha(\varphi + \pi)}{2}}_{=: \, \alpha_S(\varphi)} + 
    \underbrace{\frac{\alpha(\varphi) - \alpha(\varphi + \pi)}{2}}_{=: \, \alpha_A(\varphi)} \, .
\end{equation}
From this, we can define the symmetry coefficients
\begin{equation}
    c_{S,A} := \frac{\int_0^{2\pi} | \alpha_{S,A}(\varphi) |^2 \, \mathrm{d} \varphi}{\int_0^{2\pi} | \alpha(\varphi) |^2 \, \mathrm{d} \varphi},
\end{equation}
where $c_S$ gives the singlet component of the pair wave function and $c_A$ the triplet component. Note that we have $c_S + c_A = 1$ by construction.

\begin{figure}
    % c_{S,A}(theta)
\begin{tikzpicture}[>=latex]
    \draw (0,0) rectangle (5.0000,3.0902);
    \draw (2.5000,0) node[below]{$\theta/\pi$};
    \draw (0,1.5451) node[rotate=90, above]{$c_{S,A}(\theta)$};
    \foreach \x in {0.2000, 1.1200, 2.0400, 2.9600, 3.8800, 4.8000}{%
        \draw (\x, 0) -- ++(0, 0.1000);
        \draw (\x, 3.0902) -- ++(0, -0.1000);
    }
    \foreach \y in {0.2000, 1.5451, 2.8902}{%
        \draw (0, \y) -- ++(0.1000, 0);
        \draw (5.0000, \y) -- ++(-0.1000, 0);
    }
    \draw (0.2000, 0) node[below, scale=.8]{$0.0$};
    \draw (4.8000, 0) node[below, scale=.8]{$0.5$};
    \draw (0, 0.2000) node[left, scale=.8]{$0.0$};
    \draw (0, 2.8902) node[left, scale=.8]{$1.0$};
    \draw[thick, black, dashed] plot[smooth, shift={(0.2000, 0.2000)}] coordinates {%
        (0.0000, 0.0000) (0.0307, 0.0002) (0.0613, 0.0010) (0.0920, 0.0022) (0.1227, 0.0039) (0.1533, 0.0060) (0.1840, 0.0087) (0.2147, 0.0118) (0.2453, 0.0154) (0.2760, 0.0195) (0.3067, 0.0241) (0.3373, 0.0292) (0.3680, 0.0347) (0.3987, 0.0407) (0.4293, 0.0472) (0.4600, 0.0541) (0.4907, 0.0616) (0.5213, 0.0694) (0.5520, 0.0778) (0.5827, 0.0866) (0.6133, 0.0959) (0.6440, 0.1057) (0.6747, 0.1159) (0.7053, 0.1265) (0.7360, 0.1377) (0.7667, 0.1492) (0.7973, 0.1612) (0.8280, 0.1737) (0.8587, 0.1866) (0.8893, 0.1999) (0.9200, 0.2137) (0.9507, 0.2279) (0.9813, 0.2425) (1.0120, 0.2575) (1.0427, 0.2730) (1.0733, 0.2889) (1.1040, 0.3051) (1.1347, 0.3218) (1.1653, 0.3389) (1.1960, 0.3563) (1.2267, 0.3742) (1.2573, 0.3924) (1.2880, 0.4110) (1.3187, 0.4299) (1.3493, 0.4492) (1.3800, 0.4689) (1.4107, 0.4889) (1.4413, 0.5093) (1.4720, 0.5299) (1.5027, 0.5509) (1.5333, 0.5722) (1.5640, 0.5938) (1.5947, 0.6157) (1.6253, 0.6379) (1.6560, 0.6603) (1.6867, 0.6830) (1.7173, 0.7060) (1.7480, 0.7292) (1.7787, 0.7527) (1.8093, 0.7763) (1.8400, 0.8002) (1.8707, 0.8243) (1.9013, 0.8486) (1.9320, 0.8731) (1.9627, 0.8977) (1.9933, 0.9225) (2.0240, 0.9474) (2.0547, 0.9724) (2.0853, 0.9976) (2.1160, 1.0229) (2.1467, 1.0482) (2.1773, 1.0737) (2.2080, 1.0992) (2.2387, 1.1248) (2.2693, 1.1504) (2.3000, 1.1760) (2.3307, 1.2016) (2.3613, 1.2273) (2.3920, 1.2529) (2.4227, 1.2785) (2.4533, 1.3041) (2.4840, 1.3296) (2.5147, 1.3550) (2.5453, 1.3804) (2.5760, 1.4057) (2.6067, 1.4309) (2.6373, 1.4559) (2.6680, 1.4808) (2.6987, 1.5056) (2.7293, 1.5303) (2.7600, 1.5548) (2.7907, 1.5791) (2.8213, 1.6033) (2.8520, 1.6272) (2.8827, 1.6510) (2.9133, 1.6746) (2.9440, 1.6979) (2.9747, 1.7211) (3.0053, 1.7440) (3.0360, 1.7667) (3.0667, 1.7892) (3.0973, 1.8115) (3.1280, 1.8335) (3.1587, 1.8553) (3.1893, 1.8769) (3.2200, 1.8982) (3.2507, 1.9193) (3.2813, 1.9403) (3.3120, 1.9609) (3.3427, 1.9814) (3.3733, 2.0017) (3.4040, 2.0218) (3.4347, 2.0417) (3.4653, 2.0615) (3.4960, 2.0811) (3.5267, 2.1006) (3.5573, 2.1200) (3.5880, 2.1392) (3.6187, 2.1584) (3.6493, 2.1775) (3.6800, 2.1966) (3.7107, 2.2157) (3.7413, 2.2348) (3.7720, 2.2539) (3.8027, 2.2730) (3.8333, 2.2923) (3.8640, 2.3116) (3.8947, 2.3310) (3.9253, 2.3505) (3.9560, 2.3701) (3.9867, 2.3899) (4.0173, 2.4098) (4.0480, 2.4297) (4.0787, 2.4498) (4.1093, 2.4698) (4.1400, 2.4898) (4.1707, 2.5097) (4.2013, 2.5294) (4.2320, 2.5487) (4.2627, 2.5676) (4.2933, 2.5859) (4.3240, 2.6033) (4.3547, 2.6198) (4.3853, 2.6350) (4.4160, 2.6488) (4.4467, 2.6609) (4.4773, 2.6711) (4.5080, 2.6793) (4.5387, 2.6853) (4.5693, 2.6890) (4.6000, 2.6902)
    };
    \draw[thick, black] plot[smooth, shift={(0.2000, 0.2000)}] coordinates {%
        (0.0000, 2.6902) (0.0307, 2.6899) (0.0613, 2.6892) (0.0920, 2.6880) (0.1227, 2.6863) (0.1533, 2.6841) (0.1840, 2.6815) (0.2147, 2.6784) (0.2453, 2.6747) (0.2760, 2.6706) (0.3067, 2.6661) (0.3373, 2.6610) (0.3680, 2.6555) (0.3987, 2.6495) (0.4293, 2.6430) (0.4600, 2.6360) (0.4907, 2.6286) (0.5213, 2.6207) (0.5520, 2.6124) (0.5827, 2.6035) (0.6133, 2.5943) (0.6440, 2.5845) (0.6747, 2.5743) (0.7053, 2.5636) (0.7360, 2.5525) (0.7667, 2.5410) (0.7973, 2.5289) (0.8280, 2.5165) (0.8587, 2.5036) (0.8893, 2.4903) (0.9200, 2.4765) (0.9507, 2.4623) (0.9813, 2.4477) (1.0120, 2.4326) (1.0427, 2.4172) (1.0733, 2.4013) (1.1040, 2.3850) (1.1347, 2.3684) (1.1653, 2.3513) (1.1960, 2.3338) (1.2267, 2.3160) (1.2573, 2.2978) (1.2880, 2.2792) (1.3187, 2.2602) (1.3493, 2.2409) (1.3800, 2.2213) (1.4107, 2.2013) (1.4413, 2.1809) (1.4720, 2.1603) (1.5027, 2.1393) (1.5333, 2.1180) (1.5640, 2.0964) (1.5947, 2.0745) (1.6253, 2.0523) (1.6560, 2.0299) (1.6867, 2.0071) (1.7173, 1.9842) (1.7480, 1.9610) (1.7787, 1.9375) (1.8093, 1.9138) (1.8400, 1.8899) (1.8707, 1.8659) (1.9013, 1.8416) (1.9320, 1.8171) (1.9627, 1.7925) (1.9933, 1.7677) (2.0240, 1.7428) (2.0547, 1.7177) (2.0853, 1.6926) (2.1160, 1.6673) (2.1467, 1.6419) (2.1773, 1.6165) (2.2080, 1.5910) (2.2387, 1.5654) (2.2693, 1.5398) (2.3000, 1.5142) (2.3307, 1.4885) (2.3613, 1.4629) (2.3920, 1.4373) (2.4227, 1.4117) (2.4533, 1.3861) (2.4840, 1.3606) (2.5147, 1.3351) (2.5453, 1.3098) (2.5760, 1.2845) (2.6067, 1.2593) (2.6373, 1.2343) (2.6680, 1.2093) (2.6987, 1.1845) (2.7293, 1.1599) (2.7600, 1.1354) (2.7907, 1.1111) (2.8213, 1.0869) (2.8520, 1.0629) (2.8827, 1.0392) (2.9133, 1.0156) (2.9440, 0.9922) (2.9747, 0.9691) (3.0053, 0.9461) (3.0360, 0.9234) (3.0667, 0.9009) (3.0973, 0.8787) (3.1280, 0.8567) (3.1587, 0.8349) (3.1893, 0.8133) (3.2200, 0.7919) (3.2507, 0.7708) (3.2813, 0.7499) (3.3120, 0.7292) (3.3427, 0.7088) (3.3733, 0.6885) (3.4040, 0.6684) (3.4347, 0.6484) (3.4653, 0.6287) (3.4960, 0.6091) (3.5267, 0.5896) (3.5573, 0.5702) (3.5880, 0.5510) (3.6187, 0.5318) (3.6493, 0.5126) (3.6800, 0.4936) (3.7107, 0.4745) (3.7413, 0.4554) (3.7720, 0.4363) (3.8027, 0.4171) (3.8333, 0.3979) (3.8640, 0.3786) (3.8947, 0.3592) (3.9253, 0.3397) (3.9560, 0.3200) (3.9867, 0.3003) (4.0173, 0.2804) (4.0480, 0.2604) (4.0787, 0.2404) (4.1093, 0.2204) (4.1400, 0.2004) (4.1707, 0.1805) (4.2013, 0.1608) (4.2320, 0.1415) (4.2627, 0.1226) (4.2933, 0.1043) (4.3240, 0.0869) (4.3547, 0.0704) (4.3853, 0.0552) (4.4160, 0.0414) (4.4467, 0.0293) (4.4773, 0.0190) (4.5080, 0.0108) (4.5387, 0.0049) (4.5693, 0.0012) (4.6000, 0.0000)
    };
\end{tikzpicture}
    \caption{The  coefficients $c_S$ (solid) and $c_A$ (dashed) giving the symmetric (singlet) and antisymmetric (triplet) components of the interlayer pair
    wave function respectively, as a function of $\theta$ for $k_F\lambda =1$. }
    \label{fig:symmetry}
\end{figure}
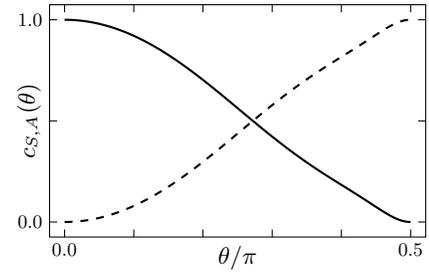

In Fig.\  \ref{fig:symmetry}, we plot these symmetry coefficients for the interlayer pairing as a function of $\theta$ for 
the layer distance $k_F\lambda =1$. For $\theta=0$ where the interaction is inversion symmetric, $c_S=1$ and $c_A=0$
since the pair wave function is purely in the singlet channel. Both $c_S$ and $c_A$ are non-zero for
$0<\theta<\pi/2$ reflecting the mixed parity pairing. The triplet component $c_A$ 
increases with increasing $\theta$ whereas $c_S$ decreases, and the pair wave function is purely in the triplet channel with $c_S=0$ and 
$c_A=1$ for $\theta=\pi/2$, where inversion symmetry is recovered.

\section{Discussion}
Dipole-dipole interactions have been shown within Hartree-Fock theory to give rise to a elliptic Fermi surface for $\theta>0$~\cite{Miyakawa2008}, and we 
should in principle include this in our analysis~\cite{Braunlich2014}. 
However, since the  symmetry of the interaction is independent of such effects, the main results of this paper, the mixed parity interlayer 
pairing for $0<\theta<\pi/2$ remain valid when Hartree-Fock effects are included. In particular, there will only be small quantitive 
changes to the phase diagram, especially for weak coupling. 

 Although our  results were derived using a weak coupling approach, we expect them to hold also for stronger coupling, 
 since they follow directly from the  symmetry of the interaction.  As   the superfluid critical temperature increases with the coupling strength, 
 this opens up the possibility  to experimentally unambiguously observe mixed parity pairing with dipolar gases. 
  Impressive experimental progress in trapping and cooling fermionic molecules with a large dipole moment have recently been reported~\cite{Heo2012,Park2015}.
  The  symmetry of the pair wave function can be detected in time-of-flight (TOF) experiments, 
  as has been explicitly demonstrated for the intralayer triplet pairing~\cite{Wu2015}.
   TOF experiments have been successfully used to detect various correlation functions in atomic gases~\cite{Greiner2005,Folling2005,Rom2006}.

The interaction cannot be made too strong however, since the system is then predicted to enter a striped (charge-density-wave) 
phase~\cite{Yamaguchi2010,Babadi2011,Block2012,Parish2012,vanZyl2015,Block2014,Keles2015,Wu2016}.
We expect the stripes to suppress the $s$-wave pairing. On the other hand, $p$-wave pairing can co-exist with the striped phase, since it is mainly formed 
in regions left ungapped by the stripes~\cite{Wu2015}. This leads to the intriguing possibility of forming a supersolid with interlayer pairing and intralayer stripes.

     Mixed parity pairing  is predicted to lead to several intriguing effects related to transport and topological properties
  of noncentrosymmetric crystals~\cite{Tanaka2009,Sato2009,Tada2009}. It is an interesting question whether similar effects appear for the present 
  system, where the lack of inversion symmetry appears through the interaction and not through an underlying crystal structure.

\section{Conclusions}
We demonstrated that dipolar fermions residing in two parallel layers can form interlayer pairing with 
both singlet and triplet components. This
mixed parity pairing arises because the interlayer interaction is not inversion symmetric when the dipole moments are
tilted with respect to the normal of the plane. We used an efficient and accurate eigenvalue 
method to investigate the competition between the interlayer pairing and the intralayer pairing, which is  in the triplet channel, 
and the resulting zero temperature phase diagram was calculated. We showed how the interlayer pair wave function smoothly 
changes from  singlet symmetry  for $\theta=0$ to  triplet symmetry for $\theta=\pi/2$. Our results indicate that  dipolar gases can be used to 
 unambiguously observe mixed parity pairing.

\appendix

\section{Numerical Method}\label{Numerics}
In order to examine the spectrum of the integral operator \eqref{eq:integral_operator}, we use the linear span of the $N\gg 1$ characteristic functions
\begin{equation*}
    b_n(\mathbf{p}) = \frac{1}{\mathcal{N}_n} \chi_{[\frac{2\pi}{N}(n-1), \frac{2\pi}{N} n]}(\varphi_{\mathbf{p}})
\end{equation*}
where $n=1,\ldots, N$. Here, $\varphi_{\mathbf{p}}$ denotes the polar angle corresponding to $\mathbf{p}$. With the normalization constants
{\small
\begin{equation*}
    \mathcal{N}_n^2 = \sqrt{\frac{2}{a^2 + b^2}} \int_{\frac{2\pi}{N}(n-1)}^{\frac{2\pi}{N} n} \mathrm{d}\varphi \, \sqrt{1 + \frac{a^2 - b^2}{a^2 + b^2} \cos(2 \varphi)} \, ,
\end{equation*}}%
these functions $(b_n)_{n=1}^N$ form an orthonormal basis with regard to the inner product
\begin{equation*}
    \braket{f | g} := \frac{2}{a^2 + b^2} \int_{\Omega_{a,b}} \overline{f(\mathbf{p})} g(\mathbf{p}) \, \mathrm{d} p \, ,
\end{equation*}
where $\Omega_{a,b}$ denotes the ellipse with half-axes $a$ and $b$. The eigenvalues of the matrix formed by $M_{mn} = \braket{b_m | A_{V} b_n}$ are an approximation for the desired spectrum. 
Note that $M$ is in fact hermitian. In the calcuations, we used $N=600$ functions. 

\acknowledgements
Partial financial support from the DFG grant GRK 1838 is gratefully acknowledged. GMB would like to acknowledge the support of the Villum Foundation via grant VKR023163 and ESF POLATOM network.


\begin{thebibliography}{43}%
\makeatletter
\providecommand \@ifxundefined [1]{%
 \@ifx{#1\undefined}
}%
\providecommand \@ifnum [1]{%
 \ifnum #1\expandafter \@firstoftwo
 \else \expandafter \@secondoftwo
 \fi
}%
\providecommand \@ifx [1]{%
 \ifx #1\expandafter \@firstoftwo
 \else \expandafter \@secondoftwo
 \fi
}%
\providecommand \natexlab [1]{#1}%
\providecommand \enquote  [1]{``#1''}%
\providecommand \bibnamefont  [1]{#1}%
\providecommand \bibfnamefont [1]{#1}%
\providecommand \citenamefont [1]{#1}%
\providecommand \href@noop [0]{\@secondoftwo}%
\providecommand \href [0]{\begingroup \@sanitize@url \@href}%
\providecommand \@href[1]{\@@startlink{#1}\@@href}%
\providecommand \@@href[1]{\endgroup#1\@@endlink}%
\providecommand \@sanitize@url [0]{\catcode `\\12\catcode `\$12\catcode
  `\&12\catcode `\#12\catcode `\^12\catcode `\_12\catcode `\%12\relax}%
\providecommand \@@startlink[1]{}%
\providecommand \@@endlink[0]{}%
\providecommand \url  [0]{\begingroup\@sanitize@url \@url }%
\providecommand \@url [1]{\endgroup\@href {#1}{\urlprefix }}%
\providecommand \urlprefix  [0]{URL }%
\providecommand \Eprint [0]{\href }%
\providecommand \doibase [0]{http://dx.doi.org/}%
\providecommand \selectlanguage [0]{\@gobble}%
\providecommand \bibinfo  [0]{\@secondoftwo}%
\providecommand \bibfield  [0]{\@secondoftwo}%
\providecommand \translation [1]{[#1]}%
\providecommand \BibitemOpen [0]{}%
\providecommand \bibitemStop [0]{}%
\providecommand \bibitemNoStop [0]{.\EOS\space}%
\providecommand \EOS [0]{\spacefactor3000\relax}%
\providecommand \BibitemShut  [1]{\csname bibitem#1\endcsname}%
\let\auto@bib@innerbib\@empty
%</preamble>
\bibitem [{\citenamefont {Anderson}(1959)}]{Anderson1959}%
  \BibitemOpen
  \bibfield  {author} {\bibinfo {author} {\bibfnamefont {P.}~\bibnamefont
  {Anderson}},\ }\href {\doibase
  http://dx.doi.org/10.1016/0022-3697(59)90036-8} {\bibfield  {journal}
  {\bibinfo  {journal} {Journal of Physics and Chemistry of Solids}\ }\textbf
  {\bibinfo {volume} {11}},\ \bibinfo {pages} {26 } (\bibinfo {year}
  {1959})}\BibitemShut {NoStop}%
\bibitem [{\citenamefont {Anderson}(1984)}]{Andersen1984}%
  \BibitemOpen
  \bibfield  {author} {\bibinfo {author} {\bibfnamefont {P.~W.}\ \bibnamefont
  {Anderson}},\ }\href {\doibase 10.1103/PhysRevB.30.4000} {\bibfield
  {journal} {\bibinfo  {journal} {Phys. Rev. B}\ }\textbf {\bibinfo {volume}
  {30}},\ \bibinfo {pages} {4000} (\bibinfo {year} {1984})}\BibitemShut
  {NoStop}%
\bibitem [{\citenamefont {Gor'kov}\ and\ \citenamefont
  {Rashba}(2001)}]{Gorkov2001}%
  \BibitemOpen
  \bibfield  {author} {\bibinfo {author} {\bibfnamefont {L.~P.}\ \bibnamefont
  {Gor'kov}}\ and\ \bibinfo {author} {\bibfnamefont {E.~I.}\ \bibnamefont
  {Rashba}},\ }\href {\doibase 10.1103/PhysRevLett.87.037004} {\bibfield
  {journal} {\bibinfo  {journal} {Phys. Rev. Lett.}\ }\textbf {\bibinfo
  {volume} {87}},\ \bibinfo {pages} {037004} (\bibinfo {year}
  {2001})}\BibitemShut {NoStop}%
\bibitem [{\citenamefont {Hayashi}\ \emph {et~al.}(2006)\citenamefont
  {Hayashi}, \citenamefont {Wakabayashi}, \citenamefont {Frigeri},\ and\
  \citenamefont {Sigrist}}]{Hayashi2006}%
  \BibitemOpen
  \bibfield  {author} {\bibinfo {author} {\bibfnamefont {N.}~\bibnamefont
  {Hayashi}}, \bibinfo {author} {\bibfnamefont {K.}~\bibnamefont
  {Wakabayashi}}, \bibinfo {author} {\bibfnamefont {P.~A.}\ \bibnamefont
  {Frigeri}}, \ and\ \bibinfo {author} {\bibfnamefont {M.}~\bibnamefont
  {Sigrist}},\ }\href {\doibase 10.1103/PhysRevB.73.092508} {\bibfield
  {journal} {\bibinfo  {journal} {Phys. Rev. B}\ }\textbf {\bibinfo {volume}
  {73}},\ \bibinfo {pages} {092508} (\bibinfo {year} {2006})}\BibitemShut
  {NoStop}%
\bibitem [{\citenamefont {Nishiyama}\ \emph {et~al.}(2007)\citenamefont
  {Nishiyama}, \citenamefont {Inada},\ and\ \citenamefont
  {Zheng}}]{Nishiyama2007}%
  \BibitemOpen
  \bibfield  {author} {\bibinfo {author} {\bibfnamefont {M.}~\bibnamefont
  {Nishiyama}}, \bibinfo {author} {\bibfnamefont {Y.}~\bibnamefont {Inada}}, \
  and\ \bibinfo {author} {\bibfnamefont {G.-q.}\ \bibnamefont {Zheng}},\ }\href
  {\doibase 10.1103/PhysRevLett.98.047002} {\bibfield  {journal} {\bibinfo
  {journal} {Phys. Rev. Lett.}\ }\textbf {\bibinfo {volume} {98}},\ \bibinfo
  {pages} {047002} (\bibinfo {year} {2007})}\BibitemShut {NoStop}%
\bibitem [{\citenamefont {Yuan}\ \emph {et~al.}(2006)\citenamefont {Yuan},
  \citenamefont {Agterberg}, \citenamefont {Hayashi}, \citenamefont {Badica},
  \citenamefont {Vandervelde}, \citenamefont {Togano}, \citenamefont
  {Sigrist},\ and\ \citenamefont {Salamon}}]{Yuan2006}%
  \BibitemOpen
  \bibfield  {author} {\bibinfo {author} {\bibfnamefont {H.~Q.}\ \bibnamefont
  {Yuan}}, \bibinfo {author} {\bibfnamefont {D.~F.}\ \bibnamefont {Agterberg}},
  \bibinfo {author} {\bibfnamefont {N.}~\bibnamefont {Hayashi}}, \bibinfo
  {author} {\bibfnamefont {P.}~\bibnamefont {Badica}}, \bibinfo {author}
  {\bibfnamefont {D.}~\bibnamefont {Vandervelde}}, \bibinfo {author}
  {\bibfnamefont {K.}~\bibnamefont {Togano}}, \bibinfo {author} {\bibfnamefont
  {M.}~\bibnamefont {Sigrist}}, \ and\ \bibinfo {author} {\bibfnamefont
  {M.~B.}\ \bibnamefont {Salamon}},\ }\href {\doibase
  10.1103/PhysRevLett.97.017006} {\bibfield  {journal} {\bibinfo  {journal}
  {Phys. Rev. Lett.}\ }\textbf {\bibinfo {volume} {97}},\ \bibinfo {pages}
  {017006} (\bibinfo {year} {2006})}\BibitemShut {NoStop}%
\bibitem [{\citenamefont {Bonalde}\ \emph {et~al.}(2005)\citenamefont
  {Bonalde}, \citenamefont {Br\"amer-Escamilla},\ and\ \citenamefont
  {Bauer}}]{Bonalde2005}%
  \BibitemOpen
  \bibfield  {author} {\bibinfo {author} {\bibfnamefont {I.}~\bibnamefont
  {Bonalde}}, \bibinfo {author} {\bibfnamefont {W.}~\bibnamefont
  {Br\"amer-Escamilla}}, \ and\ \bibinfo {author} {\bibfnamefont
  {E.}~\bibnamefont {Bauer}},\ }\href {\doibase 10.1103/PhysRevLett.94.207002}
  {\bibfield  {journal} {\bibinfo  {journal} {Phys. Rev. Lett.}\ }\textbf
  {\bibinfo {volume} {94}},\ \bibinfo {pages} {207002} (\bibinfo {year}
  {2005})}\BibitemShut {NoStop}%
\bibitem [{\citenamefont {Izawa}\ \emph {et~al.}(2005)\citenamefont {Izawa},
  \citenamefont {Kasahara}, \citenamefont {Matsuda}, \citenamefont {Behnia},
  \citenamefont {Yasuda}, \citenamefont {Settai},\ and\ \citenamefont
  {Onuki}}]{Izawa2005}%
  \BibitemOpen
  \bibfield  {author} {\bibinfo {author} {\bibfnamefont {K.}~\bibnamefont
  {Izawa}}, \bibinfo {author} {\bibfnamefont {Y.}~\bibnamefont {Kasahara}},
  \bibinfo {author} {\bibfnamefont {Y.}~\bibnamefont {Matsuda}}, \bibinfo
  {author} {\bibfnamefont {K.}~\bibnamefont {Behnia}}, \bibinfo {author}
  {\bibfnamefont {T.}~\bibnamefont {Yasuda}}, \bibinfo {author} {\bibfnamefont
  {R.}~\bibnamefont {Settai}}, \ and\ \bibinfo {author} {\bibfnamefont
  {Y.}~\bibnamefont {Onuki}},\ }\href {\doibase 10.1103/PhysRevLett.94.197002}
  {\bibfield  {journal} {\bibinfo  {journal} {Phys. Rev. Lett.}\ }\textbf
  {\bibinfo {volume} {94}},\ \bibinfo {pages} {197002} (\bibinfo {year}
  {2005})}\BibitemShut {NoStop}%
\bibitem [{\citenamefont {Mukuda}\ \emph {et~al.}(2009)\citenamefont {Mukuda},
  \citenamefont {Nishide}, \citenamefont {Harada}, \citenamefont {Iwasaki},
  \citenamefont {Yogi}, \citenamefont {Yashima}, \citenamefont {Kitaoka},
  \citenamefont {Tsujino}, \citenamefont {Takeuchi}, \citenamefont {Settai},
  \citenamefont {{\=O}nuki}, \citenamefont {Bauer}, \citenamefont {Itoh},\ and\
  \citenamefont {Haller}}]{Hidekazu2009}%
  \BibitemOpen
  \bibfield  {author} {\bibinfo {author} {\bibfnamefont {H.}~\bibnamefont
  {Mukuda}}, \bibinfo {author} {\bibfnamefont {S.}~\bibnamefont {Nishide}},
  \bibinfo {author} {\bibfnamefont {A.}~\bibnamefont {Harada}}, \bibinfo
  {author} {\bibfnamefont {K.}~\bibnamefont {Iwasaki}}, \bibinfo {author}
  {\bibfnamefont {M.}~\bibnamefont {Yogi}}, \bibinfo {author} {\bibfnamefont
  {M.}~\bibnamefont {Yashima}}, \bibinfo {author} {\bibfnamefont
  {Y.}~\bibnamefont {Kitaoka}}, \bibinfo {author} {\bibfnamefont
  {M.}~\bibnamefont {Tsujino}}, \bibinfo {author} {\bibfnamefont
  {T.}~\bibnamefont {Takeuchi}}, \bibinfo {author} {\bibfnamefont
  {R.}~\bibnamefont {Settai}}, \bibinfo {author} {\bibfnamefont
  {Y.}~\bibnamefont {{\=O}nuki}}, \bibinfo {author} {\bibfnamefont
  {E.}~\bibnamefont {Bauer}}, \bibinfo {author} {\bibfnamefont {K.~M.}\
  \bibnamefont {Itoh}}, \ and\ \bibinfo {author} {\bibfnamefont {E.~E.}\
  \bibnamefont {Haller}},\ }\href {\doibase 10.1143/JPSJ.78.014705} {\bibfield
  {journal} {\bibinfo  {journal} {Journal of the Physical Society of Japan}\
  }\textbf {\bibinfo {volume} {78}},\ \bibinfo {pages} {014705} (\bibinfo
  {year} {2009})},\ \Eprint
  {http://arxiv.org/abs/http://dx.doi.org/10.1143/JPSJ.78.014705}
  {http://dx.doi.org/10.1143/JPSJ.78.014705} \BibitemShut {NoStop}%
\bibitem [{\citenamefont {Fischer}(2006)}]{Fischer2006}%
  \BibitemOpen
  \bibfield  {author} {\bibinfo {author} {\bibfnamefont {U.~R.}\ \bibnamefont
  {Fischer}},\ }\href {\doibase 10.1103/PhysRevA.73.031602} {\bibfield
  {journal} {\bibinfo  {journal} {Phys. Rev. A}\ }\textbf {\bibinfo {volume}
  {73}},\ \bibinfo {pages} {031602} (\bibinfo {year} {2006})}\BibitemShut
  {NoStop}%
\bibitem [{\citenamefont {Li}\ \emph {et~al.}(2010)\citenamefont {Li},
  \citenamefont {Hwang},\ and\ \citenamefont {Das~Sarma}}]{Qiuzi2010}%
  \BibitemOpen
  \bibfield  {author} {\bibinfo {author} {\bibfnamefont {Q.}~\bibnamefont
  {Li}}, \bibinfo {author} {\bibfnamefont {E.~H.}\ \bibnamefont {Hwang}}, \
  and\ \bibinfo {author} {\bibfnamefont {S.}~\bibnamefont {Das~Sarma}},\ }\href
  {\doibase 10.1103/PhysRevB.82.235126} {\bibfield  {journal} {\bibinfo
  {journal} {Phys. Rev. B}\ }\textbf {\bibinfo {volume} {82}},\ \bibinfo
  {pages} {235126} (\bibinfo {year} {2010})}\BibitemShut {NoStop}%
\bibitem [{\citenamefont {Pikovski}\ \emph {et~al.}(2010)\citenamefont
  {Pikovski}, \citenamefont {Klawunn}, \citenamefont {Shlyapnikov},\ and\
  \citenamefont {Santos}}]{Pikovski2010}%
  \BibitemOpen
  \bibfield  {author} {\bibinfo {author} {\bibfnamefont {A.}~\bibnamefont
  {Pikovski}}, \bibinfo {author} {\bibfnamefont {M.}~\bibnamefont {Klawunn}},
  \bibinfo {author} {\bibfnamefont {G.~V.}\ \bibnamefont {Shlyapnikov}}, \ and\
  \bibinfo {author} {\bibfnamefont {L.}~\bibnamefont {Santos}},\ }\href
  {\doibase 10.1103/PhysRevLett.105.215302} {\bibfield  {journal} {\bibinfo
  {journal} {Phys. Rev. Lett.}\ }\textbf {\bibinfo {volume} {105}},\ \bibinfo
  {pages} {215302} (\bibinfo {year} {2010})}\BibitemShut {NoStop}%
\bibitem [{\citenamefont {Zinner}\ \emph {et~al.}(2012)\citenamefont {Zinner},
  \citenamefont {Wunsch}, \citenamefont {Pekker},\ and\ \citenamefont
  {Wang}}]{Zinner2012}%
  \BibitemOpen
  \bibfield  {author} {\bibinfo {author} {\bibfnamefont {N.~T.}\ \bibnamefont
  {Zinner}}, \bibinfo {author} {\bibfnamefont {B.}~\bibnamefont {Wunsch}},
  \bibinfo {author} {\bibfnamefont {D.}~\bibnamefont {Pekker}}, \ and\ \bibinfo
  {author} {\bibfnamefont {D.-W.}\ \bibnamefont {Wang}},\ }\href {\doibase
  10.1103/PhysRevA.85.013603} {\bibfield  {journal} {\bibinfo  {journal} {Phys.
  Rev. A}\ }\textbf {\bibinfo {volume} {85}},\ \bibinfo {pages} {013603}
  (\bibinfo {year} {2012})}\BibitemShut {NoStop}%
\bibitem [{\citenamefont {Baranov}\ \emph {et~al.}(2011)\citenamefont
  {Baranov}, \citenamefont {Micheli}, \citenamefont {Ronen},\ and\
  \citenamefont {Zoller}}]{Baranov2011}%
  \BibitemOpen
  \bibfield  {author} {\bibinfo {author} {\bibfnamefont {M.~A.}\ \bibnamefont
  {Baranov}}, \bibinfo {author} {\bibfnamefont {A.}~\bibnamefont {Micheli}},
  \bibinfo {author} {\bibfnamefont {S.}~\bibnamefont {Ronen}}, \ and\ \bibinfo
  {author} {\bibfnamefont {P.}~\bibnamefont {Zoller}},\ }\href {\doibase
  10.1103/PhysRevA.83.043602} {\bibfield  {journal} {\bibinfo  {journal} {Phys.
  Rev. A}\ }\textbf {\bibinfo {volume} {83}},\ \bibinfo {pages} {043602}
  (\bibinfo {year} {2011})}\BibitemShut {NoStop}%
\bibitem [{\citenamefont {{Camacho-Guardian}}\ \emph
  {et~al.}(2015)\citenamefont {{Camacho-Guardian}}, \citenamefont
  {{Dom{\'{\i}}nguez-Castro}},\ and\ \citenamefont {{Paredes}}}]{Camacho2015}%
  \BibitemOpen
  \bibfield  {author} {\bibinfo {author} {\bibfnamefont {A.}~\bibnamefont
  {{Camacho-Guardian}}}, \bibinfo {author} {\bibfnamefont {G.~A.}\ \bibnamefont
  {{Dom{\'{\i}}nguez-Castro}}}, \ and\ \bibinfo {author} {\bibfnamefont
  {R.}~\bibnamefont {{Paredes}}},\ }\href@noop {} {\bibfield  {journal}
  {\bibinfo  {journal} {ArXiv e-prints}\ } (\bibinfo {year} {2015})},\ \Eprint
  {http://arxiv.org/abs/1511.06311} {arXiv:1511.06311 [cond-mat.quant-gas]}
  \BibitemShut {NoStop}%
\bibitem [{\citenamefont {Matveeva}\ and\ \citenamefont
  {Giorgini}(2014)}]{Matveeva2014}%
  \BibitemOpen
  \bibfield  {author} {\bibinfo {author} {\bibfnamefont {N.}~\bibnamefont
  {Matveeva}}\ and\ \bibinfo {author} {\bibfnamefont {S.}~\bibnamefont
  {Giorgini}},\ }\href {\doibase 10.1103/PhysRevA.90.053620} {\bibfield
  {journal} {\bibinfo  {journal} {Phys. Rev. A}\ }\textbf {\bibinfo {volume}
  {90}},\ \bibinfo {pages} {053620} (\bibinfo {year} {2014})}\BibitemShut
  {NoStop}%
\bibitem [{\citenamefont {{Sigrist}}(2009)}]{Sigrist2009}%
  \BibitemOpen
  \bibfield  {author} {\bibinfo {author} {\bibfnamefont {M.}~\bibnamefont
  {{Sigrist}}},\ }in\ \href {\doibase 10.1063/1.3225489} {\emph {\bibinfo
  {booktitle} {American Institute of Physics Conference Series}}},\ \bibinfo
  {series} {American Institute of Physics Conference Series}, Vol.\ \bibinfo
  {volume} {1162},\ \bibinfo {editor} {edited by\ \bibinfo {editor}
  {\bibfnamefont {A.}~\bibnamefont {{Avella}}}\ and\ \bibinfo {editor}
  {\bibfnamefont {F.}~\bibnamefont {{Mancini}}}}\ (\bibinfo {year} {2009})\
  pp.\ \bibinfo {pages} {55--96}\BibitemShut {NoStop}%
\bibitem [{\citenamefont {Bruun}\ and\ \citenamefont
  {Taylor}(2008)}]{Bruun2008}%
  \BibitemOpen
  \bibfield  {author} {\bibinfo {author} {\bibfnamefont {G.~M.}\ \bibnamefont
  {Bruun}}\ and\ \bibinfo {author} {\bibfnamefont {E.}~\bibnamefont {Taylor}},\
  }\href {\doibase 10.1103/PhysRevLett.101.245301} {\bibfield  {journal}
  {\bibinfo  {journal} {Phys. Rev. Lett.}\ }\textbf {\bibinfo {volume} {101}},\
  \bibinfo {pages} {245301} (\bibinfo {year} {2008})}\BibitemShut {NoStop}%
\bibitem [{\citenamefont {Sieberer}\ and\ \citenamefont
  {Baranov}(2011)}]{Sieberer2011}%
  \BibitemOpen
  \bibfield  {author} {\bibinfo {author} {\bibfnamefont {L.~M.}\ \bibnamefont
  {Sieberer}}\ and\ \bibinfo {author} {\bibfnamefont {M.~A.}\ \bibnamefont
  {Baranov}},\ }\href {\doibase 10.1103/PhysRevA.84.063633} {\bibfield
  {journal} {\bibinfo  {journal} {Phys. Rev. A}\ }\textbf {\bibinfo {volume}
  {84}},\ \bibinfo {pages} {063633} (\bibinfo {year} {2011})}\BibitemShut
  {NoStop}%
\bibitem [{\citenamefont {Wu}\ \emph {et~al.}(2015)\citenamefont {Wu},
  \citenamefont {Block},\ and\ \citenamefont {Bruun}}]{Wu2015}%
  \BibitemOpen
  \bibfield  {author} {\bibinfo {author} {\bibfnamefont {Z.}~\bibnamefont
  {Wu}}, \bibinfo {author} {\bibfnamefont {J.~K.}\ \bibnamefont {Block}}, \
  and\ \bibinfo {author} {\bibfnamefont {G.~M.}\ \bibnamefont {Bruun}},\ }\href
  {\doibase 10.1103/PhysRevB.91.224504} {\bibfield  {journal} {\bibinfo
  {journal} {Phys. Rev. B}\ }\textbf {\bibinfo {volume} {91}},\ \bibinfo
  {pages} {224504} (\bibinfo {year} {2015})}\BibitemShut {NoStop}%
\bibitem [{\citenamefont {Hainzl}\ \emph {et~al.}(2008)\citenamefont {Hainzl},
  \citenamefont {Hamza}, \citenamefont {Seiringer},\ and\ \citenamefont
  {Solovej}}]{HHSS}%
  \BibitemOpen
  \bibfield  {author} {\bibinfo {author} {\bibfnamefont {C.}~\bibnamefont
  {Hainzl}}, \bibinfo {author} {\bibfnamefont {E.}~\bibnamefont {Hamza}},
  \bibinfo {author} {\bibfnamefont {R.}~\bibnamefont {Seiringer}}, \ and\
  \bibinfo {author} {\bibfnamefont {J.~P.}\ \bibnamefont {Solovej}},\ }\href
  {\doibase 10.1007/s00220-008-0489-2} {\bibfield  {journal} {\bibinfo
  {journal} {Comm. Math. Phys.}\ }\textbf {\bibinfo {volume} {281}},\ \bibinfo
  {pages} {349} (\bibinfo {year} {2008})}\BibitemShut {NoStop}%
\bibitem [{\citenamefont {Frank}\ \emph {et~al.}(2007)\citenamefont {Frank},
  \citenamefont {Hainzl}, \citenamefont {Naboko},\ and\ \citenamefont
  {Seiringer}}]{FHNS}%
  \BibitemOpen
  \bibfield  {author} {\bibinfo {author} {\bibfnamefont {R.~L.}\ \bibnamefont
  {Frank}}, \bibinfo {author} {\bibfnamefont {C.}~\bibnamefont {Hainzl}},
  \bibinfo {author} {\bibfnamefont {S.}~\bibnamefont {Naboko}}, \ and\ \bibinfo
  {author} {\bibfnamefont {R.}~\bibnamefont {Seiringer}},\ }\href {\doibase
  10.1007/BF02937429} {\bibfield  {journal} {\bibinfo  {journal} {J. Geom.
  Anal.}\ }\textbf {\bibinfo {volume} {17}},\ \bibinfo {pages} {559} (\bibinfo
  {year} {2007})}\BibitemShut {NoStop}%
\bibitem [{\citenamefont {Hainzl}\ and\ \citenamefont {Seiringer}(2008)}]{HS}%
  \BibitemOpen
  \bibfield  {author} {\bibinfo {author} {\bibfnamefont {C.}~\bibnamefont
  {Hainzl}}\ and\ \bibinfo {author} {\bibfnamefont {R.}~\bibnamefont
  {Seiringer}},\ }\href {\doibase 10.1103/PhysRevB.77.184517} {\bibfield
  {journal} {\bibinfo  {journal} {Phys. Rev. B}\ }\textbf {\bibinfo {volume}
  {77}},\ \bibinfo {pages} {184517} (\bibinfo {year} {2008})}\BibitemShut
  {NoStop}%
\bibitem [{\citenamefont {Hainzl}\ and\ \citenamefont
  {Seiringer}(2010)}]{HaSe}%
  \BibitemOpen
  \bibfield  {author} {\bibinfo {author} {\bibfnamefont {C.}~\bibnamefont
  {Hainzl}}\ and\ \bibinfo {author} {\bibfnamefont {R.}~\bibnamefont
  {Seiringer}},\ }\href {\doibase 10.1002/mana.200810195} {\bibfield  {journal}
  {\bibinfo  {journal} {Math. Nachr.}\ }\textbf {\bibinfo {volume} {283}},\
  \bibinfo {pages} {489} (\bibinfo {year} {2010})}\BibitemShut {NoStop}%
\bibitem [{\citenamefont {Hainzl}\ and\ \citenamefont
  {Seiringer}(2015)}]{CompReview}%
  \BibitemOpen
  \bibfield  {author} {\bibinfo {author} {\bibfnamefont {C.}~\bibnamefont
  {Hainzl}}\ and\ \bibinfo {author} {\bibfnamefont {R.}~\bibnamefont
  {Seiringer}},\ }\href@noop {} {\  (\bibinfo {year} {2015})},\ \Eprint
  {http://arxiv.org/abs/1511.01995v1} {arXiv:1511.01995v1} \BibitemShut
  {NoStop}%
\bibitem [{\citenamefont {Miyakawa}\ \emph {et~al.}(2008)\citenamefont
  {Miyakawa}, \citenamefont {Sogo},\ and\ \citenamefont {Pu}}]{Miyakawa2008}%
  \BibitemOpen
  \bibfield  {author} {\bibinfo {author} {\bibfnamefont {T.}~\bibnamefont
  {Miyakawa}}, \bibinfo {author} {\bibfnamefont {T.}~\bibnamefont {Sogo}}, \
  and\ \bibinfo {author} {\bibfnamefont {H.}~\bibnamefont {Pu}},\ }\href
  {\doibase 10.1103/PhysRevA.77.061603} {\bibfield  {journal} {\bibinfo
  {journal} {Phys. Rev. A}\ }\textbf {\bibinfo {volume} {77}},\ \bibinfo
  {pages} {061603} (\bibinfo {year} {2008})}\BibitemShut {NoStop}%
\bibitem [{\citenamefont {Br{\"a}unlich}\ \emph {et~al.}(2014)\citenamefont
  {Br{\"a}unlich}, \citenamefont {Hainzl},\ and\ \citenamefont
  {Seiringer}}]{Braunlich2014}%
  \BibitemOpen
  \bibfield  {author} {\bibinfo {author} {\bibfnamefont {G.}~\bibnamefont
  {Br{\"a}unlich}}, \bibinfo {author} {\bibfnamefont {C.}~\bibnamefont
  {Hainzl}}, \ and\ \bibinfo {author} {\bibfnamefont {R.}~\bibnamefont
  {Seiringer}},\ }\href {\doibase 10.1142/S0129055X14500123} {\bibfield
  {journal} {\bibinfo  {journal} {Reviews in Mathematical Physics}\ }\textbf
  {\bibinfo {volume} {26}},\ \bibinfo {pages} {1450012} (\bibinfo {year}
  {2014})},\ \Eprint
  {http://arxiv.org/abs/http://www.worldscientific.com/doi/pdf/10.1142/S0129055X14500123}
  {http://www.worldscientific.com/doi/pdf/10.1142/S0129055X14500123}
  \BibitemShut {NoStop}%
\bibitem [{\citenamefont {Heo}\ \emph {et~al.}(2012)\citenamefont {Heo},
  \citenamefont {Wang}, \citenamefont {Christensen}, \citenamefont {Rvachov},
  \citenamefont {Cotta}, \citenamefont {Choi}, \citenamefont {Lee},\ and\
  \citenamefont {Ketterle}}]{Heo2012}%
  \BibitemOpen
  \bibfield  {author} {\bibinfo {author} {\bibfnamefont {M.-S.}\ \bibnamefont
  {Heo}}, \bibinfo {author} {\bibfnamefont {T.~T.}\ \bibnamefont {Wang}},
  \bibinfo {author} {\bibfnamefont {C.~A.}\ \bibnamefont {Christensen}},
  \bibinfo {author} {\bibfnamefont {T.~M.}\ \bibnamefont {Rvachov}}, \bibinfo
  {author} {\bibfnamefont {D.~A.}\ \bibnamefont {Cotta}}, \bibinfo {author}
  {\bibfnamefont {J.-H.}\ \bibnamefont {Choi}}, \bibinfo {author}
  {\bibfnamefont {Y.-R.}\ \bibnamefont {Lee}}, \ and\ \bibinfo {author}
  {\bibfnamefont {W.}~\bibnamefont {Ketterle}},\ }\href {\doibase
  10.1103/PhysRevA.86.021602} {\bibfield  {journal} {\bibinfo  {journal} {Phys.
  Rev. A}\ }\textbf {\bibinfo {volume} {86}},\ \bibinfo {pages} {021602}
  (\bibinfo {year} {2012})}\BibitemShut {NoStop}%
\bibitem [{\citenamefont {Park}\ \emph {et~al.}(2015)\citenamefont {Park},
  \citenamefont {Will},\ and\ \citenamefont {Zwierlein}}]{Park2015}%
  \BibitemOpen
  \bibfield  {author} {\bibinfo {author} {\bibfnamefont {J.~W.}\ \bibnamefont
  {Park}}, \bibinfo {author} {\bibfnamefont {S.~A.}\ \bibnamefont {Will}}, \
  and\ \bibinfo {author} {\bibfnamefont {M.~W.}\ \bibnamefont {Zwierlein}},\
  }\href {\doibase 10.1103/PhysRevLett.114.205302} {\bibfield  {journal}
  {\bibinfo  {journal} {Phys. Rev. Lett.}\ }\textbf {\bibinfo {volume} {114}},\
  \bibinfo {pages} {205302} (\bibinfo {year} {2015})}\BibitemShut {NoStop}%
\bibitem [{\citenamefont {Greiner}\ \emph {et~al.}(2005)\citenamefont
  {Greiner}, \citenamefont {Regal}, \citenamefont {Stewart},\ and\
  \citenamefont {Jin}}]{Greiner2005}%
  \BibitemOpen
  \bibfield  {author} {\bibinfo {author} {\bibfnamefont {M.}~\bibnamefont
  {Greiner}}, \bibinfo {author} {\bibfnamefont {C.~A.}\ \bibnamefont {Regal}},
  \bibinfo {author} {\bibfnamefont {J.~T.}\ \bibnamefont {Stewart}}, \ and\
  \bibinfo {author} {\bibfnamefont {D.~S.}\ \bibnamefont {Jin}},\ }\href
  {\doibase 10.1103/PhysRevLett.94.110401} {\bibfield  {journal} {\bibinfo
  {journal} {Phys. Rev. Lett.}\ }\textbf {\bibinfo {volume} {94}},\ \bibinfo
  {pages} {110401} (\bibinfo {year} {2005})}\BibitemShut {NoStop}%
\bibitem [{\citenamefont {Folling}\ \emph {et~al.}(2005)\citenamefont
  {Folling}, \citenamefont {Gerbier}, \citenamefont {Widera}, \citenamefont
  {Mandel}, \citenamefont {Gericke},\ and\ \citenamefont
  {Bloch}}]{Folling2005}%
  \BibitemOpen
  \bibfield  {author} {\bibinfo {author} {\bibfnamefont {S.}~\bibnamefont
  {Folling}}, \bibinfo {author} {\bibfnamefont {F.}~\bibnamefont {Gerbier}},
  \bibinfo {author} {\bibfnamefont {A.}~\bibnamefont {Widera}}, \bibinfo
  {author} {\bibfnamefont {O.}~\bibnamefont {Mandel}}, \bibinfo {author}
  {\bibfnamefont {T.}~\bibnamefont {Gericke}}, \ and\ \bibinfo {author}
  {\bibfnamefont {I.}~\bibnamefont {Bloch}},\ }\href
  {http://dx.doi.org/10.1038/nature03500} {\bibfield  {journal} {\bibinfo
  {journal} {Nature}\ }\textbf {\bibinfo {volume} {434}},\ \bibinfo {pages}
  {481} (\bibinfo {year} {2005})}\BibitemShut {NoStop}%
\bibitem [{\citenamefont {Rom}\ \emph {et~al.}(2006)\citenamefont {Rom},
  \citenamefont {Best}, \citenamefont {van Oosten}, \citenamefont {Schneider},
  \citenamefont {Folling}, \citenamefont {Paredes},\ and\ \citenamefont
  {Bloch}}]{Rom2006}%
  \BibitemOpen
  \bibfield  {author} {\bibinfo {author} {\bibfnamefont {T.}~\bibnamefont
  {Rom}}, \bibinfo {author} {\bibfnamefont {T.}~\bibnamefont {Best}}, \bibinfo
  {author} {\bibfnamefont {D.}~\bibnamefont {van Oosten}}, \bibinfo {author}
  {\bibfnamefont {U.}~\bibnamefont {Schneider}}, \bibinfo {author}
  {\bibfnamefont {S.}~\bibnamefont {Folling}}, \bibinfo {author} {\bibfnamefont
  {B.}~\bibnamefont {Paredes}}, \ and\ \bibinfo {author} {\bibfnamefont
  {I.}~\bibnamefont {Bloch}},\ }\href {http://dx.doi.org/10.1038/nature05319}
  {\bibfield  {journal} {\bibinfo  {journal} {Nature}\ }\textbf {\bibinfo
  {volume} {444}},\ \bibinfo {pages} {733} (\bibinfo {year}
  {2006})}\BibitemShut {NoStop}%
\bibitem [{\citenamefont {Yamaguchi}\ \emph {et~al.}(2010)\citenamefont
  {Yamaguchi}, \citenamefont {Sogo}, \citenamefont {Ito},\ and\ \citenamefont
  {Miyakawa}}]{Yamaguchi2010}%
  \BibitemOpen
  \bibfield  {author} {\bibinfo {author} {\bibfnamefont {Y.}~\bibnamefont
  {Yamaguchi}}, \bibinfo {author} {\bibfnamefont {T.}~\bibnamefont {Sogo}},
  \bibinfo {author} {\bibfnamefont {T.}~\bibnamefont {Ito}}, \ and\ \bibinfo
  {author} {\bibfnamefont {T.}~\bibnamefont {Miyakawa}},\ }\href {\doibase
  10.1103/PhysRevA.82.013643} {\bibfield  {journal} {\bibinfo  {journal} {Phys.
  Rev. A}\ }\textbf {\bibinfo {volume} {82}},\ \bibinfo {pages} {013643}
  (\bibinfo {year} {2010})}\BibitemShut {NoStop}%
\bibitem [{\citenamefont {Babadi}\ and\ \citenamefont
  {Demler}(2011)}]{Babadi2011}%
  \BibitemOpen
  \bibfield  {author} {\bibinfo {author} {\bibfnamefont {M.}~\bibnamefont
  {Babadi}}\ and\ \bibinfo {author} {\bibfnamefont {E.}~\bibnamefont
  {Demler}},\ }\href {\doibase 10.1103/PhysRevB.84.235124} {\bibfield
  {journal} {\bibinfo  {journal} {Phys. Rev. B}\ }\textbf {\bibinfo {volume}
  {84}},\ \bibinfo {pages} {235124} (\bibinfo {year} {2011})}\BibitemShut
  {NoStop}%
\bibitem [{\citenamefont {Block}\ \emph {et~al.}(2012)\citenamefont {Block},
  \citenamefont {Zinner},\ and\ \citenamefont {Bruun}}]{Block2012}%
  \BibitemOpen
  \bibfield  {author} {\bibinfo {author} {\bibfnamefont {J.~K.}\ \bibnamefont
  {Block}}, \bibinfo {author} {\bibfnamefont {N.~T.}\ \bibnamefont {Zinner}}, \
  and\ \bibinfo {author} {\bibfnamefont {G.~M.}\ \bibnamefont {Bruun}},\ }\href
  {http://stacks.iop.org/1367-2630/14/i=10/a=105006} {\bibfield  {journal}
  {\bibinfo  {journal} {New Journal of Physics}\ }\textbf {\bibinfo {volume}
  {14}},\ \bibinfo {pages} {105006} (\bibinfo {year} {2012})}\BibitemShut
  {NoStop}%
\bibitem [{\citenamefont {Parish}\ and\ \citenamefont
  {Marchetti}(2012)}]{Parish2012}%
  \BibitemOpen
  \bibfield  {author} {\bibinfo {author} {\bibfnamefont {M.~M.}\ \bibnamefont
  {Parish}}\ and\ \bibinfo {author} {\bibfnamefont {F.~M.}\ \bibnamefont
  {Marchetti}},\ }\href {\doibase 10.1103/PhysRevLett.108.145304} {\bibfield
  {journal} {\bibinfo  {journal} {Phys. Rev. Lett.}\ }\textbf {\bibinfo
  {volume} {108}},\ \bibinfo {pages} {145304} (\bibinfo {year}
  {2012})}\BibitemShut {NoStop}%
\bibitem [{\citenamefont {van Zyl}\ \emph {et~al.}(2015)\citenamefont {van
  Zyl}, \citenamefont {Kirkby},\ and\ \citenamefont {Ferguson}}]{vanZyl2015}%
  \BibitemOpen
  \bibfield  {author} {\bibinfo {author} {\bibfnamefont {B.~P.}\ \bibnamefont
  {van Zyl}}, \bibinfo {author} {\bibfnamefont {W.}~\bibnamefont {Kirkby}}, \
  and\ \bibinfo {author} {\bibfnamefont {W.}~\bibnamefont {Ferguson}},\ }\href
  {\doibase 10.1103/PhysRevA.92.023614} {\bibfield  {journal} {\bibinfo
  {journal} {Phys. Rev. A}\ }\textbf {\bibinfo {volume} {92}},\ \bibinfo
  {pages} {023614} (\bibinfo {year} {2015})}\BibitemShut {NoStop}%
\bibitem [{\citenamefont {Block}\ and\ \citenamefont
  {Bruun}(2014)}]{Block2014}%
  \BibitemOpen
  \bibfield  {author} {\bibinfo {author} {\bibfnamefont {J.~K.}\ \bibnamefont
  {Block}}\ and\ \bibinfo {author} {\bibfnamefont {G.~M.}\ \bibnamefont
  {Bruun}},\ }\href {\doibase 10.1103/PhysRevB.90.155102} {\bibfield  {journal}
  {\bibinfo  {journal} {Phys. Rev. B}\ }\textbf {\bibinfo {volume} {90}},\
  \bibinfo {pages} {155102} (\bibinfo {year} {2014})}\BibitemShut {NoStop}%
\bibitem [{\citenamefont {{Keles}}\ and\ \citenamefont
  {{Zhao}}(2015)}]{Keles2015}%
  \BibitemOpen
  \bibfield  {author} {\bibinfo {author} {\bibfnamefont {A.}~\bibnamefont
  {{Keles}}}\ and\ \bibinfo {author} {\bibfnamefont {E.}~\bibnamefont
  {{Zhao}}},\ }\href@noop {} {\bibfield  {journal} {\bibinfo  {journal} {ArXiv
  e-prints}\ } (\bibinfo {year} {2015})},\ \Eprint
  {http://arxiv.org/abs/1512.06147} {arXiv:1512.06147 [cond-mat.quant-gas]}
  \BibitemShut {NoStop}%
\bibitem [{\citenamefont {Wu}\ \emph {et~al.}(2016)\citenamefont {Wu},
  \citenamefont {Block},\ and\ \citenamefont {Bruun}}]{Wu2016}%
  \BibitemOpen
  \bibfield  {author} {\bibinfo {author} {\bibfnamefont {Z.}~\bibnamefont
  {Wu}}, \bibinfo {author} {\bibfnamefont {J.~K.}\ \bibnamefont {Block}}, \
  and\ \bibinfo {author} {\bibfnamefont {G.~M.}\ \bibnamefont {Bruun}},\ }\href
  {http://dx.doi.org/10.1038/srep19038} {\bibfield  {journal} {\bibinfo
  {journal} {Scientific Reports}\ }\textbf {\bibinfo {volume} {6}},\ \bibinfo
  {pages} {19038 EP } (\bibinfo {year} {2016})}\BibitemShut {NoStop}%
\bibitem [{\citenamefont {Tanaka}\ \emph {et~al.}(2009)\citenamefont {Tanaka},
  \citenamefont {Yokoyama}, \citenamefont {Balatsky},\ and\ \citenamefont
  {Nagaosa}}]{Tanaka2009}%
  \BibitemOpen
  \bibfield  {author} {\bibinfo {author} {\bibfnamefont {Y.}~\bibnamefont
  {Tanaka}}, \bibinfo {author} {\bibfnamefont {T.}~\bibnamefont {Yokoyama}},
  \bibinfo {author} {\bibfnamefont {A.~V.}\ \bibnamefont {Balatsky}}, \ and\
  \bibinfo {author} {\bibfnamefont {N.}~\bibnamefont {Nagaosa}},\ }\href
  {\doibase 10.1103/PhysRevB.79.060505} {\bibfield  {journal} {\bibinfo
  {journal} {Phys. Rev. B}\ }\textbf {\bibinfo {volume} {79}},\ \bibinfo
  {pages} {060505} (\bibinfo {year} {2009})}\BibitemShut {NoStop}%
\bibitem [{\citenamefont {Sato}\ and\ \citenamefont
  {Fujimoto}(2009)}]{Sato2009}%
  \BibitemOpen
  \bibfield  {author} {\bibinfo {author} {\bibfnamefont {M.}~\bibnamefont
  {Sato}}\ and\ \bibinfo {author} {\bibfnamefont {S.}~\bibnamefont
  {Fujimoto}},\ }\href {\doibase 10.1103/PhysRevB.79.094504} {\bibfield
  {journal} {\bibinfo  {journal} {Phys. Rev. B}\ }\textbf {\bibinfo {volume}
  {79}},\ \bibinfo {pages} {094504} (\bibinfo {year} {2009})}\BibitemShut
  {NoStop}%
\bibitem [{\citenamefont {{Tada}}\ \emph {et~al.}(2009)\citenamefont {{Tada}},
  \citenamefont {{Kawakami}},\ and\ \citenamefont {{Fujimoto}}}]{Tada2009}%
  \BibitemOpen
  \bibfield  {author} {\bibinfo {author} {\bibfnamefont {Y.}~\bibnamefont
  {{Tada}}}, \bibinfo {author} {\bibfnamefont {N.}~\bibnamefont {{Kawakami}}},
  \ and\ \bibinfo {author} {\bibfnamefont {S.}~\bibnamefont {{Fujimoto}}},\
  }\href {\doibase 10.1088/1367-2630/11/5/055070} {\bibfield  {journal}
  {\bibinfo  {journal} {New Journal of Physics}\ }\textbf {\bibinfo {volume}
  {11}},\ \bibinfo {eid} {055070} (\bibinfo {year} {2009})},\ \Eprint
  {http://arxiv.org/abs/0902.3043} {arXiv:0902.3043 [cond-mat.supr-con]}
  \BibitemShut {NoStop}%
\end{thebibliography}
\end{document}